\documentclass[aps,nofootinbib,reprint,amsmath,amssymb,aps,showpacs,groupedaddress,superscriptaddress,nofootinbib]{revtex4-1}
\usepackage{graphicx}% Include figure files
\usepackage{dcolumn}% Align table columns on decimal point
\usepackage{bm}
\usepackage{color}
\usepackage{soul}
 \newcommand{\comment}[1]{}

\begin{document}

\title{Dissipative dynamics in a quantum bistable system: Crossover from weak to strong damping}
\author{Luca Magazz\`u}\email{luca.magazzu@unipa.it}
\affiliation{Dipartimento di Fisica e Chimica,
Universit\`{a} di Palermo, Viale delle Scienze, Edificio 18, I-90128 Palermo, Italy}
\affiliation{Radiophysics Department, Lobachevsky State University of Nizhni Novgorod, Russia}
\author{Davide Valenti}\email{davide.valenti@unipa.it}
\affiliation{Dipartimento di Fisica e Chimica,
Universit\`{a} di Palermo, Viale delle Scienze, Edificio 18, I-90128 Palermo, Italy}
\author{Bernardo Spagnolo}\email{bernardo.spagnolo@unipa.it}
\affiliation{Dipartimento di Fisica e Chimica,
Universit\`{a} di Palermo, Viale delle Scienze, Edificio 18, I-90128 Palermo, Italy}
\affiliation{Radiophysics Department, Lobachevsky State University of Nizhni Novgorod, Russia}
\affiliation{Istituto Nazionale di Fisica Nucleare, Sezione di Catania, Italy}
\author{Milena Grifoni}\email{milena.grifoni@physik.uni-regensburg.de}
\affiliation{Theoretische Physik, Universit\"at Regensburg, 93040 Regensburg, Germany}

\date{\today}

\begin{abstract}
The dissipative dynamics of a quantum bistable system coupled to a Ohmic heat bath is investigated beyond the spin-boson approximation. Within the path-integral approach to quantum dissipation, we propose an approximation scheme which exploits the separation of time scales between intra- and interwell (tunneling) dynamics. The resulting generalized master equation for the populations in a space localized basis enables us to investigate a wide range of temperatures and system-environment coupling strengths.
A phase diagram in the coupling-temperature space is provided to give a comprehensive account of the different dynamical regimes. 
\end{abstract}

\pacs{03.65.Yz, 05.40.-a, 85.25.-j, 05.60.Gg}  

\maketitle

\section{Introduction}
\label{intro}

Many quantum systems of interest in disparate areas of physics, ranging from particle physics to condensed matter or chemical physics, are characterized by the presence of two minima of potential energy separated by a potential barrier. In the classical regime, if  the barrier crossing is not thermally induced, a non-driven particle settles indefinitely in a potential minimum. In contrast, in  the quantum regime, even at zero temperature, the particle is confined in one potential well until the escape by tunneling occurs.\\ 
\indent A prominent example is the tunneling of a flux quantum through the effective potential barrier created by a Josephson junction in a superconducting circuit~\cite{Chiorescu2003,Chiarello2012}. Other multi-state bistable systems, which received much attention as candidate for both classical and quantum computing hardware, are molecular nanomagnets (see Ref.~\cite{Gatteschi2006} and references therein). The experimental signature of thermally induced tunneling of magnetization in these high-spin molecules is the presence of resonant peaks in the transfer rate from a metastable state, as a function of an applied static bias~\cite{Friedman2010}. Coherent manipulation of the quantum state in molecular nanomagnets has been performed in Refs.~\cite{Schlegel2008,Takahashi2009}. Moreover, asymmetric bistable systems are used to investigate the relaxation from  a quantum metastable state~\cite{ThorwartPRL2000,Thorwart2001,Caldara2011,Spagnolo2012,Magazzu2013,Valenti2015}.\\ 
\indent Depending on the potential barrier, which in some cases (e.g. in superconducting devices) can be manipulated, the tunneling dynamics take  place on time scales much larger than those associated to the intrawell motion. This allows for the coherent manipulation of quantum states for quantum simulations and computation~\cite{Nakamura1999,VanderWal2000,Vion2002,Yu2002,Martinis2002,Devoret2004,Clarke2008,Ladd2010,You2011,Devoret2013}. Bistability emerges also in the dynamics of quantum dots depending on the electron-phonon interaction~\cite{Wilner2013}.
\\ 
\indent A feature which is present in every real quantum system is the coupling to a dissipative environment which causes relaxation and decoherence. It is therefore desirable to characterize the noise sources and their influence on dynamical and coherence properties for various coupling/temperature regimes~\cite{Weiss2012,Paladino2014}. For example,  the existence of quantum coherence has been demonstrated in strongly dissipative environments, such as those involved in energy transport in biological systems~\cite{Chin2013}.\\ 
\indent  In general, if the system was prepared, say, in the left well, coherent Rabi oscillations between the two metastable wells occur at very small dissipation strengths and low temperatures. On the other hand, at sufficiently large damping  and/or high temperatures the dynamics is known to be incoherent. Further, localization is predicted at large enough damping~\cite{LeHur2008}.
So far, the coherent to incoherent crossover has been only investigated in the so-called two-level system (TLS)  approximation for the Hilbert space of the bistable system: The temperature is taken to be low enough so, to a good approximation, the dynamics can be restricted to the space spanned  by the lowest doublet $\{|g\rangle,|e\rangle\}$ of eigenstates of  the bare Hamiltonian (see Fig.~\ref{fig1}).
A vast literature exists~\cite{Leggett1987,*LeggettErratum1995,Weiss2012,Paladino2014} which investigates the coherent-to-incoherent crossover for various dissipation mechanisms in  great detail.\\ 
\indent However, for temperatures of the order of the separation between the lowest and the next-lying energy levels, the TLS approximation breaks down and the multi-level nature of the bistable potential cannot be neglected.  This occurs, for example, in chemical processes where the presence of  a higher tunneling doublet under an effective potential barrier accounts for activated rates at appropriate temperatures~\cite{Cukier1990}.
 Despite its relevance for applications, the dissipative bistable dynamics in this temperature regime has been poorly investigated so far~\cite{DekkerI1991,*DekkerII1991,*DekkerIV1991, Cukier1995,Morillo1996,Cukier1997,ThorwartPRL2000,Thorwart2001}.\\ 
\indent  For very small damping strengths, a perturbative Bloch-Redfield approach capturing coherent intrawell and interwell oscillations  is appropriate~\cite{Morillo1996,Cukier1997}. In the opposite regime of moderate to large damping and temperatures the dynamics is fully incoherent and is well described in terms of rate equations for the populations of states localized in the wells, with rates obtained within a nonperturbative path-integral approach~\cite{ThorwartPRL2000}. However, the \emph{crossover}  regime, characterized by moderate damping and temperatures, presents an unsolved challenge. The main difficulty  lies in the fact that, as the tunneling dynamics occurs on a time scale much larger than that of the intrawell dynamics, a hybrid situation can occur where quantum coherence is present at the level of intrawell motion, but is lost at longer times where tunneling processes are relevant.\\ 
\indent In this work we propose a novel, nonperturbative in the coupling, approximation scheme. Within this scheme, we derive a generalized master equation which enables us to investigate the transient and long time bistable dynamics of a multi-level open system in the crossover regime. Moreover, we provide a phase diagram in the coupling-temperature space which describes the various dynamical regimes occurring at different  dissipation strengths.
\indent Similarly to Refs.~\cite{ThorwartPRL2000,Thorwart2001}, we use a real-time path-integral approach for the reduced density matrix (RDM) of a quantum particle linearly coupled to a bosonic reservoir. By tracing out the reservoir degrees of freedom, a formally exact expression for the elements of the RDM is obtained, in which the environmental effects are encapsulated in the so-called Feynman-Vernon influence functional~\cite{Feynman1963}. The latter introduces nonlocal in time correlations between paths, a feature which requires suitable approximation schemes. To take into account the different time scales of the dynamics, we include long- and short-time correlations in the intrawell motion, while only short-time correlations turn out to be relevant for the tunneling process. At low temperature and weak damping, the resulting dynamics exhibits coherent oscillations at short times and incoherent tunneling behavior at longer times. By increasing the temperature and/or coupling strength, a transition to a fully incoherent regime is observed, in accordance with the predictions in Ref.~\cite{ThorwartPRL2000,Thorwart2001}.\\
\indent The paper is organized as follows. In Sec.~\ref{model} we introduce the Caldeira-Leggett model of dissipation in open quantum systems. An overview of path-integral techniques for the dissipative TLS is given. The validity of the TLS description is discussed and path-integral approximation schemes are introduced which constitute the building blocks of developments for multi-level systems.
The specific multi-level system considered in this work, the \emph{double-doublet} system, is described in terms of the spatially localized basis of the \emph{discrete variable representation}.\\
\indent In Sec.~\ref{PI approach} we introduce the path-integral representation of the reduced dynamics whose expression in the discrete variable representation is the starting point for the approximations made in Sec.~\ref{approximations}. A generalized master equation, capturing the non-Markovian character of the reduced dynamics, is derived in Sec.~\ref{GME-vrwiba}  within the novel approximation scheme developed.\\
\indent In Sec.~\ref{phase diagram} we introduce the phase diagram, obtained by the combined use of Bloch-Redfield and path-integral techniques, which describes the various dynamical regimes of the dissipative double-doublet system. The phase diagram gives a comprehensive account of the problem of the dissipative quantum dynamics beyond the TLS approximation and, along with the novel approximation scheme, constitutes the main result of the present work. Examples of experiments on real physical systems which can be described by the model of multi-level system considered in our work are provided.\\
\indent Finally,  in Sec.~\ref{results} several examples of dynamics, at dissipation regimes ranging from weak coupling/low temperature to strong coupling/high temperature, are shown, and in Sec.~\ref{conclusions} we summarize our findings and make some  final remarks.
 \section{Model}
\label{model}
The open system is a quantum particle of mass $M$ and coordinates $\hat{q}$ and $\hat{p}$, subject to a double-well potential $V(\hat{q})$. According to the Caldeira-Leggett model~\cite{Caldeira1981}, the particle interacts linearly with a so-called bosonic heat bath, a reservoir of $N$ independent quantum harmonic oscillators of frequencies $\omega_{j}$, in the thermodynamical limit $N\rightarrow\infty$. 
 \subsection{Hamiltonian}
\label{hamiltonian}
The full Hamiltonian of  the model is
\begin{equation}\label{Hfull}
\hat{H}=\hat{H}_{S}+\hat{H}_{B}+\hat{H}_{SB},
\end{equation}
where the bare system Hamiltonian is $\hat{H}_{S}=\hat{p}^{2}/2M+V(\hat{q})$.  The double-well potential is parametrized by the quartic function~\cite{ThorwartPRL2000}
\begin{equation}\label{Potential}
V(\hat{q})=\frac{M^2\omega^4_0}{64\Delta
U}\hat{q}^4-\frac{M\omega^2_0}{4}\hat{q}^2-\epsilon\hat{q},
\end{equation}
where $\epsilon$ is the asymmetry and $\Delta U$ the barrier height at $\epsilon=0$.
In Eq.~(\ref{Hfull}) $\hat{H}_{B}$ and $\hat{H}_{SB}$ describe, according to the Caldeira-Leggett model,  the free bath energy and the particle-bath interaction energy, respectively. Their contribution is given by
\[
%\begin{equation}\label{HSB}
\hat{H}_{B}+\hat{H}_{SB}=\frac{1}{2}\sum_{j=1}^{N} \left[
\frac{\hat{p}^{2}_{j}}{m_{j}}+m_{j}\omega^{2}_{j}\left ( \hat{x}_{j}
-\frac{c_{j}} {m_{j}\omega^{2}_{j} } \hat{q} \right )^{2}\right].
%\end{equation}
\]
The interaction term features the bilinear coupling $c_{j}\hat{x}_{j}\hat{q}$ and a renormalization term $\propto \hat{q}^{2}$. The latter compensates for the contribution of the oscillators in the effective potential \emph{felt} by the particle, thus giving a purely dissipative bath.\\ 
\indent  The bath spectral density function is defined by  
\begin{equation}\label{J}
J(\omega)=\frac{\pi}{2}\sum_{j=1}^{N}\frac{c_{j}^{2}}{m_{j}\omega_{j}}\delta(\omega-\omega_{j}).
\end{equation}
 In the continuum limit $N\rightarrow \infty$, i.e., in the presence of a large broadband reservoir, $J$ is phenomenologically modeled as $J\propto\omega^{s}$, with a cutoff at high frequency~\cite{Weiss2012}.\\
\indent Throughout this work we consider spectral density functions of the form
\begin{equation}\label{J-continuous}
J(\omega)=M\gamma_{s}\omega_{ph}^{1-s}\omega^{s}\exp(-\omega/\omega_{c}),
\end{equation}
where $\omega_{c}$ is the cutoff frequency and $\omega_{ph}$ is a  characteristic frequency scale of the heat bath. For $s<1$ the bath is said \emph{sub-Ohmic}, for $s=1$ \emph{Ohmic}, and for $s>1$ \emph{super-Ohmic}.\\
\indent  In the quantum Langevin equation for the model with Ohmic dissipation ($s=1$) the parameter $\gamma_{1}\equiv \gamma$ is the frequency-independent damping constant. Correspondingly, the Langevin  equation features a memoryless damping kernel and therefore, in the classical limit, the Ohmic bath reduces to a white noise source. By comparing the discrete and continuous versions of $J$, one recognizes that $\gamma$ is a measure of the overall system-bath coupling. For this reason we refer to $\gamma$ as \emph{coupling strength}.
\subsection{The two-level system}
\label{TLS}
\begin{figure}[htbp]
\begin{center}
\includegraphics[width=7.5cm,angle=0]{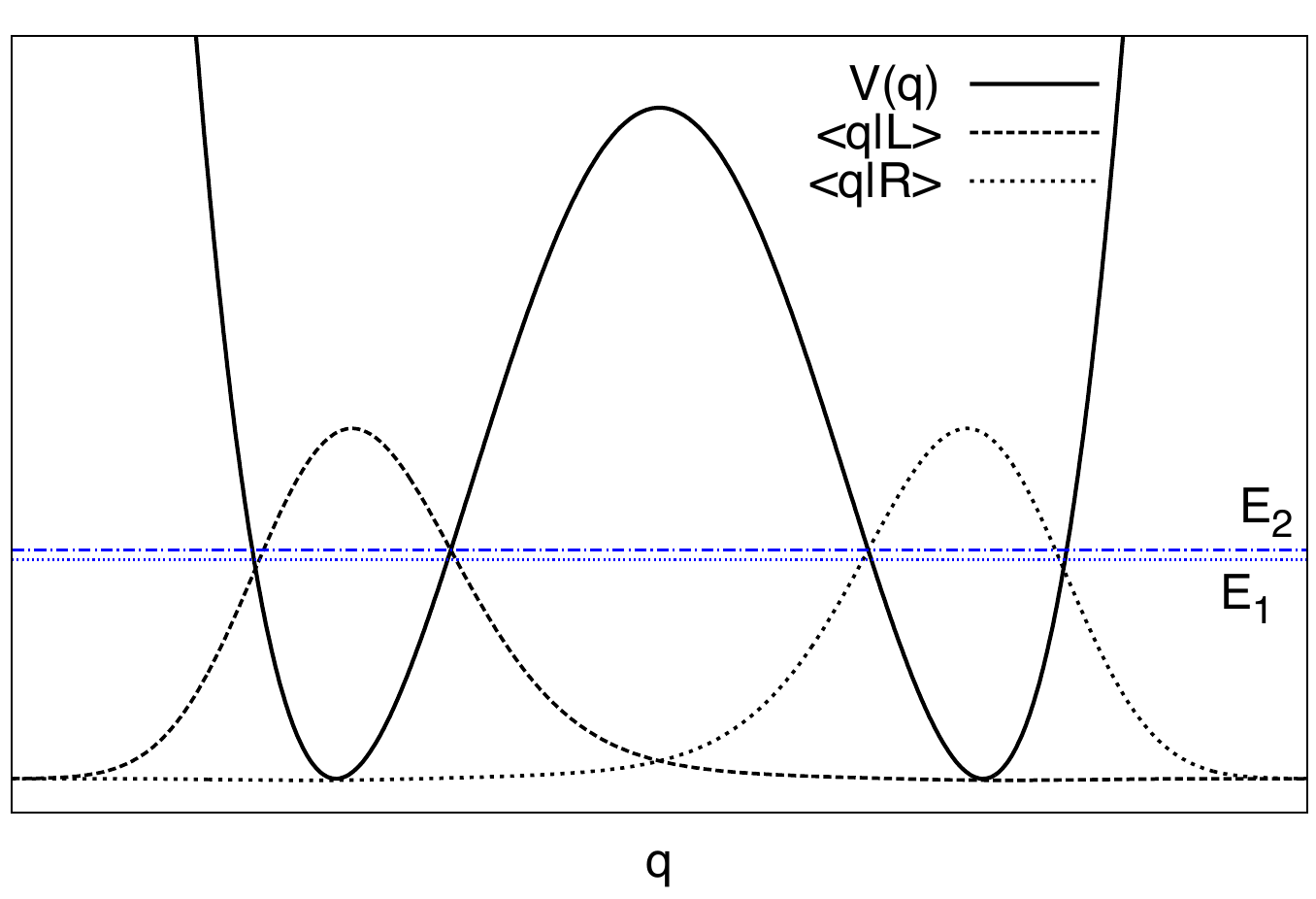}\\
\caption{\small{(Color online) Symmetric double-well potential  ($\epsilon=0$ in Eqs.~(\ref{Potential}) and~(\ref{HTLS})), first two energy levels (horizontal lines)  and localized states $|L/R\rangle=(|E_{1}\rangle\mp|E_{2}\rangle)/\sqrt{2}$.}}
\label{fig1}
\end{center}
\end{figure}

If  the particle is initially in a superposition of the two lower energy states ($|E_{1}\rangle,|E_{2}\rangle$) of the potential $V$, and the temperature is low enough, to a good approximation  the system can be considered a two-level system (TLS) and the model in Eq.~(\ref{Hfull}) reduces to the celebrated \emph{spin-boson} model~\cite{Leggett1987}. A picture of the TLS dynamics in terms of  tunneling from one well to the other is given by the localized basis $\{|R\rangle,|L\rangle\}$ depicted in Fig.~\ref{fig1}. In this basis  the free TLS Hamiltonian reads 
\begin{equation}\label{HTLS}
\hat{H}_{TLS}=-\frac{\hbar}{2}(\Delta \sigma_{x}+\epsilon\sigma_{z}),
 \end{equation}
where $\Delta$ is the tunneling frequency and $\epsilon$ is the bias. The spin operators in the localized basis read $\sigma_z=|R\rangle\langle R|-|L\rangle\langle L|$ and $\sigma_x=|R\rangle\langle L|+|L\rangle\langle R|$.\\
\indent For the spin-boson dynamics there exist various approximation schemes.
When the system is weakly coupled to the heat bath (usually this is the case for quantum optical systems and qubit setups), the approach traditionally used is based on the Born-Markov master equation for the reduced density operator~\cite{Blum2012}. This approach captures well the coherent tunneling dynamics, characterized by the relaxation and dephasing rates $\Gamma_{rel}=\tau_{1}^{-1}$ and $\Gamma_{ph}=\tau_{2}^{-1}$, respectively.
 However, its perturbative in the coupling character makes the Born-Markov master equation approach unsuited for situations where the coupling is not weak. In these cases real-time path-integral techniques can be used to trace out the bath degrees of freedom and obtain a still exact formal expression for the reduced dynamics. In some cases, this expression can be numerically evaluated by tensor multiplication~\cite{Makri1995} or using Monte Carlo or stochastic techniques~\cite{EggerPRB1994,Stockburger2002,Koch2008,Orth2013}. However, the numerical evaluation of the path-integral is a hard task, especially at long times. It is therefore convenient to have an equation of the Nakajima-Zwanzig  type~\cite{Petruccione2002} which captures the non-Markovian character of the reduced dynamics. Yet, also in this case difficulties arise  in obtaining a reasonably simple expression for the kernels. To overcome the problem different approximation schemes exist,within the real-time path-integral approach,  yielding  integro-differential master equations for the populations in a spatially localized representation.\\
\indent In the path-integral representation for open systems the trace operation on the  bath degrees of freedom yield a factorized form for the amplitude associated to a path: the bare amplitude, relative to the free system, is multiplied by the Feynman-Vernon  influence functional~\cite{Feynman1963} which weights the amplitude according to the effect exerted by the particle's  motion on the bath.\\ 
\indent  The influence functional introduces time nonlocal correlations in the amplitudes, making the path-integral expression intractable for anharmonic potentials. For the spin-boson dynamics the simplest approximation, nonperturbative in the coupling, is the noninteracting blip approximation (NIBA)~\cite{Leggett1987}. The NIBA neglects nonlocal correlations at high temperature and/or strong coupling. This approach, while  being nonperturbative in the coupling, is perturbative in the tunneling element $\Delta$ (see Eq.~(\ref{HTLS})).\\ 
\indent  In the opposite regime, the weak coupling approximation~\cite{Gorlich1989}, which treats the coupling to the first order and $\Delta$ to all orders, is appropriate. It gives the same results as the Born-Markov approach. Finally, an approach exists which interpolates between these two extrema by considering the local correlations fully and the nonlocal ones to the first order in the coupling. This scheme is called weakly interacting blip approximation (WIBA)~\cite{NesiPRB2007}, and, by construction, also covers the intricate regime of intermediate temperatures and damping, where both the weak coupling and the NIBA fail.\\ 
\indent   However, beyond the TLS approximation, this crossover regime is not accessible to currently existing  Bloch-Redfield-like~\cite{Morillo1996,Cukier1997} or NIBA-like~\cite{ThorwartPRL2000,Thorwart2001} approximation schemes.\\ 
\indent  In this work we consider a generalization of the spin-boson model to a four-level bistable system. Exploiting the difference in time scales between the fast intrawell motion and the slow interwell (tunneling) dynamics, we are able to treat the first dynamics according to the WIBA scheme and the second one according to the NIBA. The resulting scheme covers the crossover region of intermediate temperatures and coupling strengths.
\subsection{The double-doublet system}
\label{DDS}
Including successive energy states beyond the first two, a localized basis can still be constructed, as for the TLS (see Fig.~\ref{fig1}). Suppose that the Hilbert space of the system is spanned by the first $M$ energy states $|E_{1}\rangle,\dots,|E_{M}\rangle$. We pass to the so-called \emph{discrete variable representation} (DVR)~\cite{Harris1965} by performing a unitary transformation $T$ which diagonalizes the position operator $\hat{q}$ in this truncated Hilbert space: 
\begin{equation}
\begin{aligned}
\textbf{q}^{DVR}=&\textbf{T}\textbf{q}\textbf{T}^{\dag}\\
=&\text{diag}\{Q_{1},\dots,Q_{M}\},
\end{aligned}
\end{equation}
where $\textbf{q}$ is the matrix representing $\hat{q}$ in the energy basis.
In the DVR the basis states $|Q_{i}\rangle$ are eigenstates of $\hat{q}$ \emph{localized} around the $M$ eigenvalues $Q_{1},\dots,Q_{M}$ and are  related to the energy eigenbasis by 
\begin{equation}\label{energy-DVR}
|Q_{j}\rangle=\sum_{k=1}^{M}T^{*}_{jk}|E_{k}\rangle,
\end{equation}
where $T_{ij}=(\textbf{T})_{ij}$.
Note that the spatial discretization resulting from the transformation in the reduced Hilbert space reflects the approximate treatment of putting an upper limit ($E_{M}$) on the available energies, namely of considering the system a $M$-level system: A spatially continuous picture is recovered for $M\rightarrow \infty$.\\
 \begin{figure}[htbp]
\begin{center}
\includegraphics[width=7.5cm,angle=0]{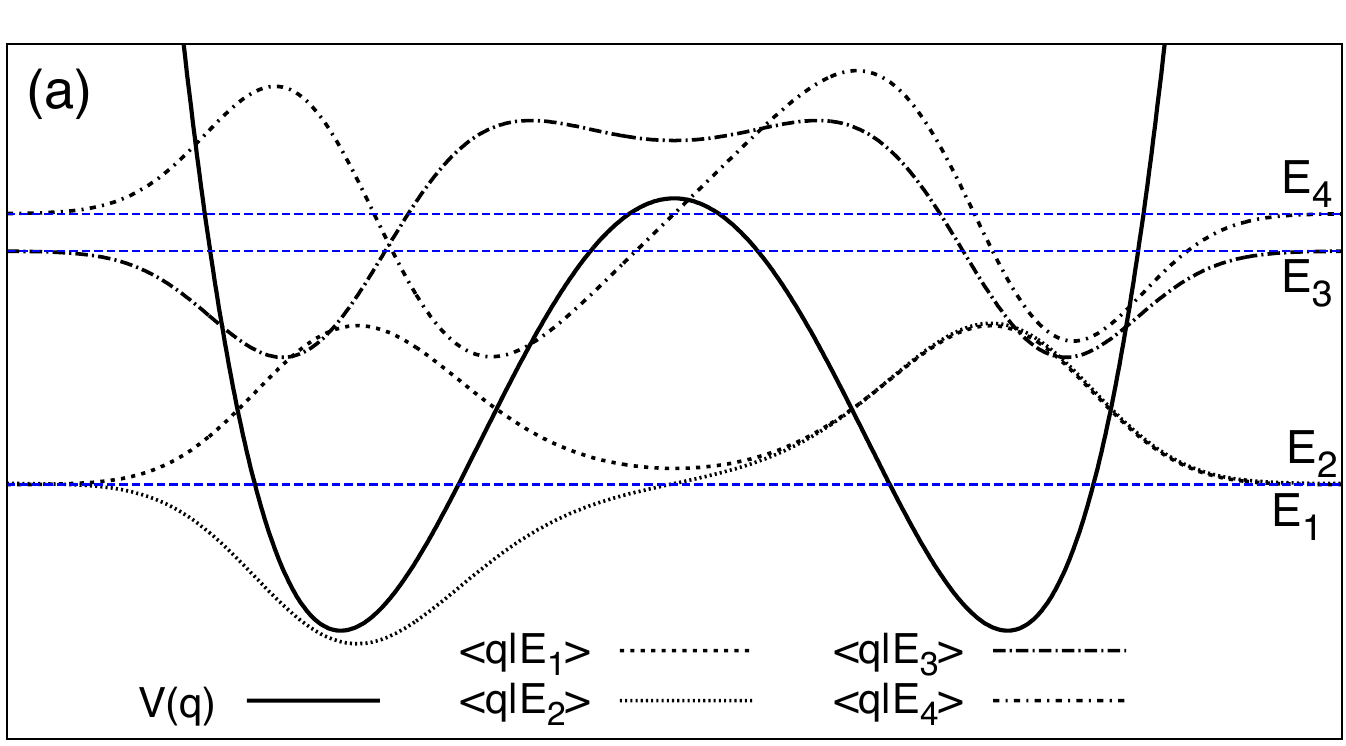}
\includegraphics[width=7.5cm,angle=0]{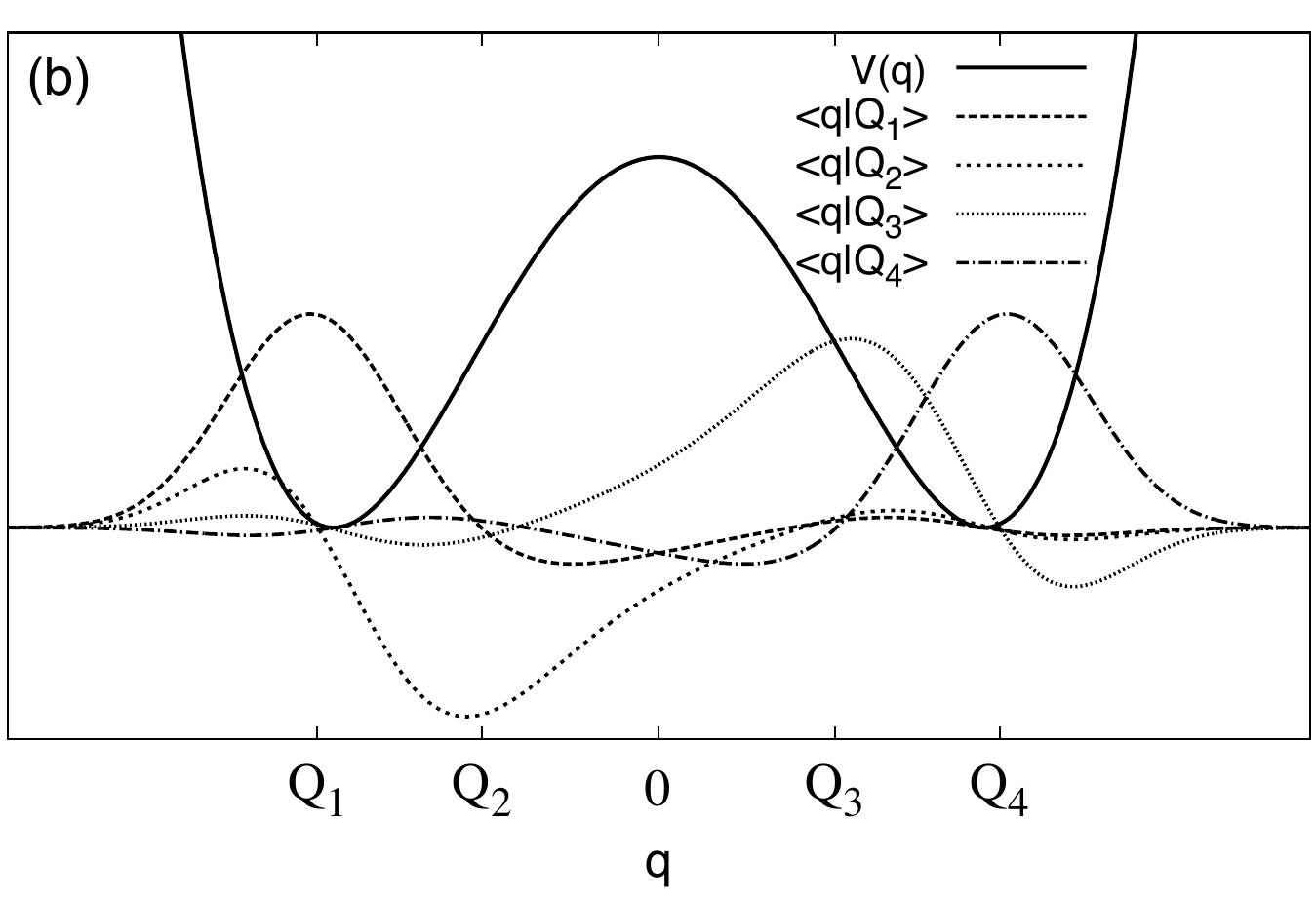}
\caption{\small{(Color online) Potential $V(q)$ (Eq.~(\ref{Potential})) with $\Delta U=1.4~\hbar\omega_{0}$ and $\epsilon=0$. The minima are at $q\simeq\mp 3.35 \sqrt{\hbar/(M\omega_{0})}$. ($a$) - The first four energy levels (horizontal lines) and the respective energy eigenfunctions.  ($b$) - Discrete variable representation. The four positions $Q_{j}$ and the corresponding basis functions $\langle q|Q_{j}\rangle$ (see Eq.~(\ref{energy-DVR})). With the chosen parameters for $V$ we get $Q_{1/4}\simeq\mp 3.51\sqrt{\hbar/(M\omega_{0})}$ and $Q_{2/3}\simeq\mp 1.82\sqrt{\hbar/(M\omega_{0}).}$}}
\label{fig2}
\end{center}
\end{figure}
\indent The DVR was developed in the context of molecular quantum dynamics, where one deals with multidimensional problems with possibly complicated potential surfaces~\cite{Light2000}. One can pass to the DVR starting from finite orthogonal basis set other than the (generally unknown) energy eigenbasis. In this case the DVR involves an approximate evaluation of the Hamiltonian matrix elements and other possible sources of inaccuracies  (not present in our treatment), which are less severe as the number $M$ of basis states is increased~\cite{Light2000}.\\
\indent However, our use of the DVR is not aimed at solving the Schr\"odinger equation for the \emph{isolated} system, which can be done numerically (see  Fig.~\ref{fig2}(a)) with a little effort. The motivation is in fact that the Feynman-Vernon influence functional, a feature of the path-integral treatment of the open dynamics, is most conveniently expressed and approximated using  this spatially localized representation for the particle (see Secs.~\ref{exact RDM} and~\ref{approximations}).\\ 
\indent In Fig.~\ref{fig2} the energy and the DVR eigenstates  are shown for the specific four-state system considered in this work, i.e., the symmetric double-well potential ($\epsilon=0$) with the relevant Hilbert space spanned by the first $M=4$ energy eigenstates. Since the corresponding energy levels are arranged in a pair of two well separated doublets, the system is called  \emph{double-doublet system}.\\
\indent The energy doublets are characterized by internal frequency differences
$\Omega_{2}= \omega_{2}-\omega_{1}\simeq 0.0037\omega_{0}$
and $\Omega_{1}=  \omega_{4}- \omega_{3}\simeq 0.1212 \omega_{0}\gg\Omega_{2}$, where $\omega_{i}=E_{i}/\hbar$.
The other relevant frequency of the problem is the average inter-doublet frequency spacing
\begin{equation}\label{Omega0}
\Omega_{0} =\frac{ \omega_{4}+ \omega_{3}}{2}-\frac{ \omega_{2}+ \omega_{1}}{2}\simeq0.8151\omega_{0}.
\end{equation}
As a consequence, the following inequalities hold for the characteristic frequencies of our system
\begin{equation}\label{Inequality}
\Omega_{0}>\Omega_{1}\gg\Omega_{2}.
 \end{equation}
In terms of the inter-doublet frequency spacing $\Omega_{0}$, the validity of the double-doublet approximation is given by $T\lesssim\hbar\Omega_0/k_{B}$.\\
\indent The DVR basis is composed by the four functions peaked around the positions $Q_{1},\dots, Q_{4}$, as shown in Fig.~\ref{fig2}(b). Their decomposition in terms of  energy eigenstates is
\begin{equation}\label{dvr-basis}
\begin{aligned}
&|Q_{1/4}\rangle=\frac{v}{\sqrt{2}}\Big(|E_{1}\rangle\mp|E_{2}\rangle-u|E_{3}\rangle\pm u|E_{4}\rangle\Big),\\
&|Q_{2/3}\rangle=\frac{v}{\sqrt{2}}\Big(\mp u|E_{1}\rangle+u|E_{2}\rangle\mp |E_{3}\rangle+|E_{4}\rangle\Big),
\end{aligned}
\end{equation}
where $v=(1+u^{2})^{-1/2}$ and $u\simeq 0.585$.\\ 
\indent  Throughout this work we consider the populations of the double-doublet system in the DVR
\begin{equation}
\rho_{ii}=\langle Q_{i}|\rho|Q_{i}\rangle \qquad\quad (i=1,\dots,4),
\end{equation}
being the probabilities to find the particle in regions centered at the eigenvalues $Q_{i}$.\\ 
\indent  In Fig.~\ref{fig3} the free dynamics of the system is shown with initial condition
\begin{equation}\label{free-DDS-init-cond}
\rho(0)=|Q_{1}\rangle\langle Q_{1}|.
\end{equation}
A comparison is made with the free evolution of the same system in the TLS approximation with initial condition $\rho(0)=|L\rangle\langle L|$ (see Fig.~\ref{fig1}).\\
\vspace{0.3cm}
 \begin{figure}[htbp]
\begin{center}
\includegraphics[width=8cm,angle=0]{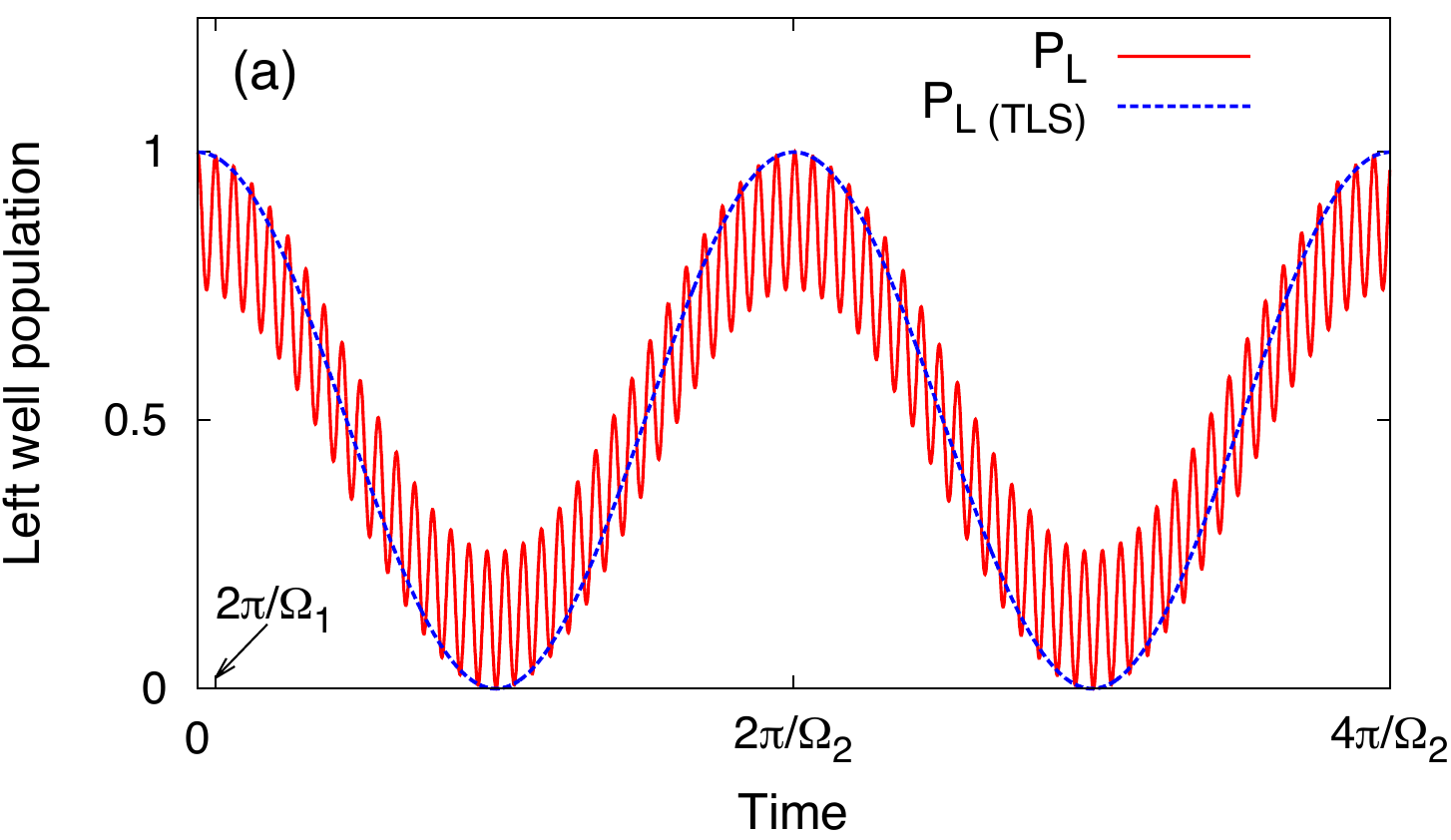}
\includegraphics[width=8cm,angle=0]{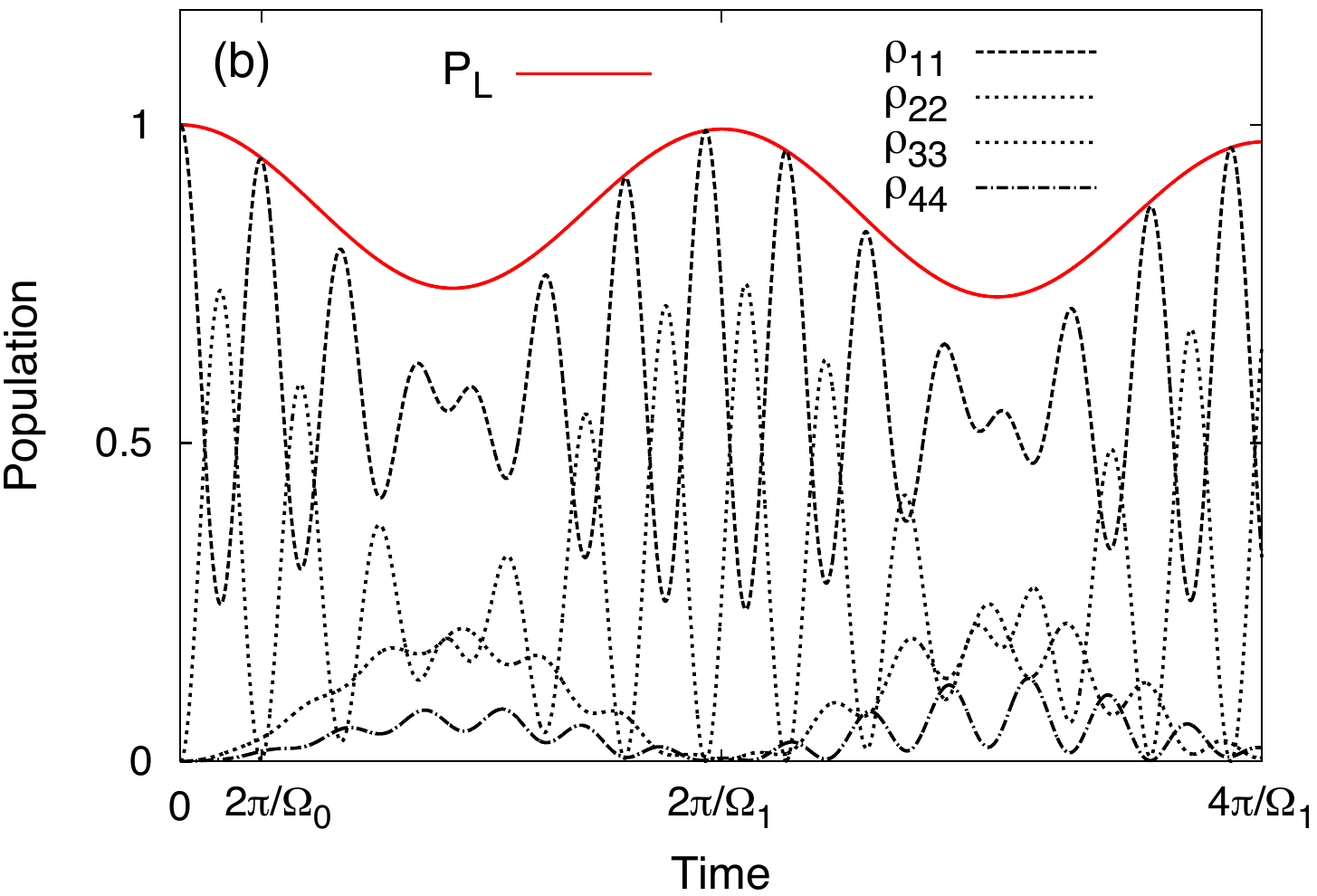}
\caption{\small{(Color online) Free system initially in the state $|Q_1\rangle$. ($a$) - left-well population $P_L=\rho_{11}+\rho_{22}$ vs time. Comparison with the time evolution of the left-well population for the system in the two-level system approximation (TLS) initially in the state $|L\rangle$. ($b$) - left-well population $P_{L}$ and individual populations vs time. Frequencies $\Omega_{i}$ are defined in Sec.~\ref{DDS}.}}
\label{fig3}
\end{center}
\end{figure}

\indent Initial condition~(\ref{free-DDS-init-cond}) involves all of the four energy states, thus the resulting dynamics comprises the three relevant time scales of the problem.
Specifically, the time evolution of the free system displays fast intrawell oscillations, at frequency $\Omega_{0}$, and a tunneling dynamics occurring at two distinct time scales: The shorter is given by $\Omega_{1}$ and the longer occurs on the times scale set by $\Omega_{2}$.
As shown in Fig.~\ref{fig3}, the long time oscillations coincide with those of  the TLS starting in the state $|L\rangle$.
This rich dynamical behavior reflects the configuration of the energy levels in the present problem. The presence of different time scales in the free dynamics  will be exploited in Sec.~\ref{approximations} to make approximations on the Feynman-Vernon influence functional in the presence of dissipation.\\
\indent  Before going into the details of the path-integral approach to dissipative dynamics, it is appropriate to discuss the role of the initial condition in determining, along with the dissipation regime, the validity of the TLS approximation. This is done in the next section. 
\subsection{Initial preparation and validity of the two-level system approximation}
\label{free DDS}
\begin{figure}[htbp]
\begin{center}
\includegraphics[width=8cm,angle=0]{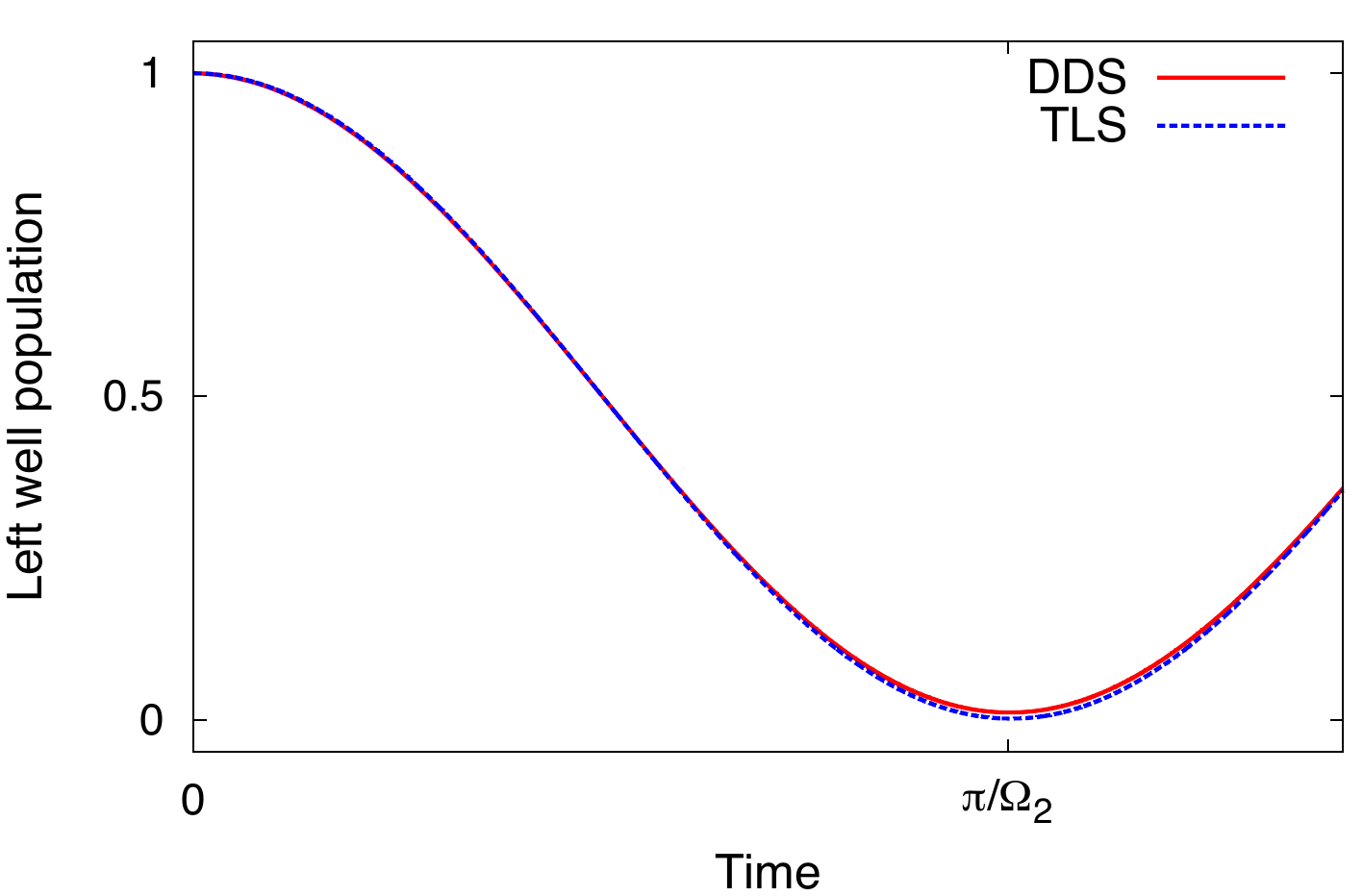}
\caption{\small{(Color online) Time evolution of the left-well population  $P_{L}=\rho_{11}+\rho_{22}$ of the double-doublet system (DDS) initially in the state $|L\rangle=v\left(|Q_{1}\rangle-u|Q_{2}\rangle\right)$. Comparison with the left-well population vs time for the system in the TLS approximation. The frequency $\Omega_{2}$ is the lower doublet frequency spacing (see Sec.~\ref{DDS}). $\gamma=0.0001~\omega_{0}$ and $T=0.0001~\hbar\omega_{0}/k_{B}$.}}
\label{fig4}
\end{center}
\end{figure}

At low temperature ($T\ll \hbar\Omega_{0}/k_{B}$) the TLS approximation is appropriate and yields the same predictions as those obtained considering higher energy states, provided that no energy levels other than the first two are involved in the initial preparation\footnote{This is not generally true for a driven system.}. \\ 
\indent  This is exemplified in Fig.~\ref{fig4} where, at weak coupling and low temperature, the time evolution of the left state of the TLS is compared with the left-well population $P_{L}=\rho_{11}+\rho_{22}$ of the double-doublet system initially prepared in the state $|L\rangle$ which, in the DVR basis~(\ref{dvr-basis}), reads
\begin{equation}
\label{init-cond-DDS2}
|L\rangle=v\left(|Q_{1}\rangle-u|Q_{2}\rangle\right).\\
\end{equation}

\vspace{0.3cm}
In the dissipation regime considered in Fig.~\ref{fig4}, our system is well described by the technique of the Bloch-Redfield master equation~\cite{Blum2012} 
 \begin{equation}\label{BR_ME-2}
\dot{\rho}_{nm}^{E}(t)=-i\omega_{nm}\rho_{nm}^{E}(t)+\sum_{k,l}\mathcal{L}_{nm,kl}\rho_{kl}^{E}(t)
\end{equation}
(the superscript $E$ indicates the energy representation).
This (Born-Markov approximated) master equation is derived from the microscopic model introduced in Sec.~\ref{hamiltonian} and is characterized by the so-called dephasing rates $|\mathcal{L}_{ij,ij}|$, fixing the coherence time of a  superposition of energy states $E_{i}$ and $E_{j}$. As a consequence $|\mathcal{L}_{ij,ij}|$ also fixes the time scale of decay of an oscillatory behavior of frequency $\omega_{ij}=\omega_{i}-\omega_{j}$ in the space localized picture given by the discrete variable representation. The analytical solution of Eq.~(\ref{BR_ME-2}) is discussed in Appendix~\ref{bloch-redfield}.\\
\indent  As expected in this weak coupling/low temperature regime, the Bloch-Redfield approach predicts the same behavior for double-doublet system with initial condition~(\ref{init-cond-DDS2}) and for the TLS initially in the state $|L\rangle$ (see Fig.~\ref{fig4}). Notice that no oscillations of frequency $\Omega_{1}$ are present in the time evolution of $P_{L}$.\\ 
\indent  Due to their perturbative character, the Born-Markov approaches~\cite{DekkerI1991,*DekkerII1991,*DekkerIV1991} fail in the intermediate to strong coupling regime and nonperturbative techniques are needed, such as those based on path-integrals which we introduce in Sec.~\ref{PI approach}. There, along the line of Ref.~\cite{Thorwart2001}, a formally exact expression for the reduced density matrix of our multi-level system is derived in the localized basis given in Eq.~(\ref{dvr-basis}). Approximations which go beyond the NIBA-like generalization to multi-level systems proposed in Ref.~\cite{Thorwart2001} are discussed in Sec.~\ref{approximations}.  Notice that the parameters of the bistable potential are chosen as in Ref.~\cite{Thorwart2001}, such that some of the results presented there for the incoherent regime can be used as reference. 
\section{Path-integral approach}
\label{PI approach}
The path-integral approach, being inherently nonperturbative in the coupling, has been proven useful to treat the strong as well as the weak dissipation regime~\cite{Weiss2012}. Within this approach the dynamical object of interest is the reduced density matrix in the position representation
\begin{equation}\label{RDM-a}
\rho_{qq'}(t)=\langle q|Tr _{B}\mathcal{W}(t)|q'\rangle,
\end{equation}
where $\mathcal{W}$ is the full system plus reservoir density operator and $Tr_{B}$ denotes the trace over the bath degrees of freedom. The time evolution of $\mathcal{W}$ is induced by the operator $U(t,t_{0})=\exp(-i\hat{H} (t-t_{0})/\hbar )$, with $\hat{H}$ the full Hamiltonian of the model (see Eq.~(\ref{Hfull})), according to
\[
\mathcal{W}(t)=U(t,t_{0})\mathcal{W}(t_{0})U^{\dag}(t,t_{0}).
\]
We assume the following factorized initial condition
\begin{equation}\label{init-cond}
\mathcal{W}(t_{0})=\rho(t_{0})\otimes R^{Th},
\end{equation}
 with the particle in the state $\rho(t_{0})$ and the bath in the canonical thermal equilibrium state $R^{Th}=Z^{-1}\exp(-\beta\hat{H}_{B})$.\\
\indent The reduced density matrix at time $t_{f}$ is then given by
\begin{equation} \label{RDM}
\begin{aligned}
\rho_{qq'}(t_{f})=\int dq_0\int dq'_0 G(q,q',t_{f};q_0,q'_0,t_{0})\rho_{q_{0}q_{0}'}(t_{0}).
\end{aligned}
\end{equation}
The propagator for the density matrix has the form of a double path-integral over the left ($q$) and right ($q'$) coordinates 
\begin{equation}
\begin{aligned}\label{PIpropagator}
G&(q,q',t_{f};q_0,q'_0,t_{0})\\
=&\int^{q}_{q_0}\mathcal{D}q(t)\int^{q'}_{q'_0}\mathcal{D}^*q'(t) e^{\frac{i}{\hbar}(S[q(t)]-S[q'(t)])}\mathcal{F}_{FV}[q(t),q'(t)].
\end{aligned} 
\end{equation}
In this path-integral expression, the action functional $S[q(t)]$ for the bare system is
\begin{equation}
S[q(t)]=\int_{t_0}^{t}dt'\left(\frac{p^{2}(t')}{2M}-V(q(t')) \right).
\end{equation}
The Feynman-Vernon (FV) influence functional $\mathcal{F}_{FV}$, which couples the paths $q$ and $q'$, derives from tracing over the bath degrees of freedom and is equal to $1$ for vanishing coupling. This in turn results in the factorization of the two path-integrals in Eq.~(\ref{PIpropagator}) for the isolated system.\\
\indent Denoting by $\mathbf{x}$ the collective position coordinate of the bath  oscillators, the FV influence functional reads
\begin{equation}\label{FVgeneral}
\begin{aligned}
\mathcal{F}_{FV}&[q(t),q'(t)]= \int  d\mathbf{x}_0  d\mathbf{x'}_0 R_{\mathbf{x}_0\mathbf{x'}_0}^{Th}\int d\mathbf{x}_{f}\\ 
&\times \int^{\mathbf{x}_{f}}_{\mathbf{x}_0}\mathcal{D}\mathbf{x}(t) \int^{\mathbf{x}_{f}}_{\mathbf{x}_0'}\mathcal{D}^*\mathbf{x'}(t)e^{\frac{i}{\hbar}\left( S[q(t),\mathbf{x}(t)]-S[q'(t),\mathbf{x}'(t)]\right)}.
\end{aligned} 
\end{equation}
The action functional $S[q(t),\mathbf{x}(t)]$ of the bath oscillators subject to the influence of the particle's motion is
\[
\begin{aligned}
S&[q(t),\mathbf{x}(t)]\\
=&\frac{1}{2}\sum_{j=1}^{N}\int_{t_0}^{t_{f}}dt'\left[
\frac{p^{2}_{j}(t')}{m_{j}}-m_{j}\omega^{2}_{j}\left ( x_{j}(t')
-\frac{c_{j}} {m_{j}\omega^{2}_{j} } q(t') \right )^{2}\right]. 
\end{aligned}
\]
The path-integrals for the bath oscillators in Eq.~(\ref{FVgeneral}) can be solved analytically, being a set of Gaussian integrals and yield the following exact expression
\begin{equation}\label{FV}
\mathcal{F}_{FV}=\exp(-\Phi_{FV}),
\end{equation}
where the influence phase functional $\Phi_{FV}$ takes the form
\begin{equation}\label{phiFV}
\begin{aligned}
\Phi_{FV}[y(t),x(t)]=&\frac{1}{\hbar^{2}}\int_{t_0}^{t_{f}}dt'\int_{t_0}^{t'}dt''y(t')\\
&\times \left[L'(t'-t'')y(t'')+iL''(t'-t'')x(t'') \right]\\
&+i\frac{\mu}{2\hbar^{2}}\int_{t_0}^{t_{f}}dt' x(t')y(t').
\end{aligned}
\end{equation}
Here we have introduced the relative and center of mass coordinates 
\begin{equation}\label{yx}
y=q-q' \quad \text{and} \quad x=q+q',
\end{equation}
the bath force correlation function
\begin{equation}\label{L}
\begin{aligned}
L(t)=&L'(t)+iL''(t)\\
=&\frac{\hbar}{\pi}\int^{\infty}_0 d\omega J(\omega)\left( \coth{\frac{\hbar\omega\beta}{2}}\cos{\omega t}-i\sin{\omega t}\right),
\end{aligned}
\end{equation}
and $\mu=2\hbar/\pi\int^{\infty}_0 d\omega J(\omega)/\omega$~\cite{Weiss2012}.\\
\indent If the particle is free or in harmonic potentials, the path-integral in Eq.~(\ref{PIpropagator}) can be evaluated analytically~\cite{Grabert1988}. In our case, the nonlinearity of the double-well potential does not allow for an exact evaluation of $\rho_{qq'}(t)$ and approximations are required. The first approximation is the truncation of the Hilbert space to the first few energy states leading to the  spatially discretized picture described in Sec.~\ref{DDS}, the discrete variable representation. In the next section we give the path-integral expression of the propagator for the reduced density matrix in this representation. This will be the basis for further approximations on the influence functional discussed in Sec.~\ref{approximations}.
\subsection{Propagator in the discrete variable representation}
\label{exact RDM}
\begin{figure}[htbp]
\includegraphics[width=8cm,angle=0]{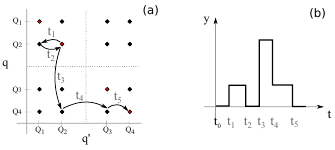}
\caption{\small{(Color online) ($a$) - Example of a double path in the plane $(q,q')$ in which the two diagonal sites $(Q_{2},Q_{2})$ and $(Q_{4},Q_{4})$ are connected by five transitions. When the path crosses the horizontal (vertical) dotted line, the coordinate $q$ ($q'$) makes a tunneling transition. ($b$) - Time resolved picture in terms of  the relative coordinate $y=q-q'$.}}
\label{fig5}
\end{figure}

In the spatially discretized picture discussed in Sec.~\ref{DDS}, a path $(q,q')$ of the reduced density matrix is no more a smooth function of the time but a walk on the two-dimensional spatial grid with grid points $\{Q_{1},\dots,Q_{4}\}$. An example of path with five transitions is shown in Fig.~\ref{fig5}.
In the discrete variable representation the path-integral turns into a sum over all the possible path configurations with time integrations over the transition times.
 The discretized version of Eq.~(\ref{RDM}) reads
\begin{equation}\label{RDM-DVR}
\rho_{qq'}(t_{f})=\sum_{q_{0},q'_{0}=Q_1}^{Q_4}G(q,q',t_{f};q_{0},q'_{0},t_{0})\rho_{q_{0}q'_{0}}(t_{0}),
\end{equation}
where the propagator $G$ is
\begin{equation}\label{propagator}
\begin{aligned}
G(q,q',&t_{f};q_{0},q'_{0},t_{0})\\
=&\sum_{n=0}^{\infty}\int_{t_{0}}^{t_{f}}D_{n}\{t_{j}\}\mathcal{A}[q]\mathcal{A}^{*}[q']\mathcal{F}_{FV}[x,y].
\end{aligned}
\end{equation}
The sum is over the number $n$ of transitions and the symbol $\int_{t_0}^{t}D_{n}\{t_{j}\}$ denotes the following sum of integrals over the transition times
\begin{equation}\label{int-D}
\int_{t_{0}}^{t_{f}}D_{n}\{t_{j}\}=\sum_{\text{paths}_{n}}\int_{t_0}^{t_{f}}dt_{n}\int_{t_0}^{t_{n}}dt_{n-1}\dots\int_{t_0}^{t_2}dt_{1}.
\end{equation}
The subscript $\text{\emph{paths}}_{n}$ stands for the set of all possible path configurations with $n$ transitions at times $t_{j}$ with $j=1,\dots,n$.
  The bare amplitude for one of these paths is 
\begin{equation}\label{BareAmp}
\begin{aligned}
\mathcal{A}[q]\mathcal{A}^{*}[q']=B_{0}(t_{1}-t_{0})\prod_{j=1}^{n}\left(-i\right)\Delta_j B_{j}(t_{j+1}-t_j),
\end{aligned}
\end{equation}
where $t_{n+1}\equiv t_{f}$. The transition amplitudes per unit time $\Delta_j$ are defined by
\begin{equation}\label{delta-j}
\Delta_j=
\left\{ 
  \begin{array}{l l}
 \frac{1}{\hbar} \langle q_{j}|\hat{H}_{S}| q_{j-1}\rangle   & \quad \text{for a $q$ transition}\\
  -\frac{1}{\hbar} \langle q'_{j}|\hat{H}_{S}| q'_{j-1}\rangle & \quad \text{for a $q'$ transition},
  \end{array} \right.
  \end{equation}
where $q/q'_{j-1}$ and $q/q'_{j}$ are the values assumed by the coordinates $q/q'$ at times $t_{j-1}$ and $t_{j}$, respectively. At each time the position coordinates take on values in the set $\{Q_{1},\dots,Q_{4}\}$. 
The \emph{bias factors} $B_{j}$ read
\begin{equation}
B_{j}(t_{j+1}-t_j)=\exp\left[-i\epsilon_{j}(t_{j+1}-t_{j})\right],
\end{equation}
where 
\begin{equation}\label{epsilon-j}
\epsilon_{j}=\frac{1}{\hbar}\left(\langle q_{j}|\hat{H}_{S}| q_{j}\rangle-\langle q'_{j}|\hat{H}_{S}| q'_{j}\rangle\right).
\end{equation}
The factors $\Delta_{j}$ and $\epsilon_{j}$ are the multi-state generalizations of tunneling element and bias introduced in the TLS Hamiltonian (Eq. (\ref{HTLS})). For the sake of readability, from now on we use two indexes $i,j=1,\dots,4$ for the above-mentioned factors, that is, $\Delta_{ij}$ and $\epsilon_{ij}$, to specify the values $Q_{i}$ and $Q_{j}$ assumed by the position coordinates in Eqs.~(\ref{delta-j}) and~(\ref{epsilon-j}). 
 Numerical values of $\Delta_{ij}$ and $\epsilon_{ij}$ for the present problem are given in Appendix~\ref{parameters}.\\ 
\indent  In the discrete variable representation, the phase of the Feynman-Vernon influence functional (Eq.~(\ref{FV})) reads~\cite{Weiss2012}
\begin{equation}\label{FV-DVR}
\Phi[\xi,\chi]_{FV}=-\sum_{i=1}^{n}\sum_{j=0}^{i-1}\left(\xi_{i}S_{ij}\xi_{j}+i\xi_{i}R_{ij}\chi_{j}\right),
\end{equation}
where, in terms of the center of mass ($x$) and relative ($y$) coordinates, the so-called \emph{charges} are defined by
\begin{equation}\label{charges}
\begin{aligned}
\xi_{j}&=y_{j}-y_{j-1},\\
\chi_{j}&=x_{j}-x_{j-1}.
\end{aligned}
\end{equation}
In Eq.~(\ref{FV-DVR}) we used the shorthand notation $S_{ij}/R_{ij}:=S/R(t_{i}-t_{j})$.
The function $Q(t)=S(t)+iR(t)$ is called \emph{pair interaction}, as it couples $\xi$- and $\chi$-charges in the time nonlocal fashion described by Eq.~(\ref{FV-DVR}).
The pair interaction $Q(t)$ is related to the bath force correlation function $L(t)$, defined in Eq.~(\ref{L}), by 
\begin{equation}
L(t)=\hbar^{2}\frac{d^{2}Q(t)}{dt^{2}}.
\end{equation}  
\indent  In what follows we use the Ohmic spectral density function with exponential cutoff $J(\omega)=M\gamma\omega\exp(-\omega/\omega_{c})$, where $\omega_{c}\gg \omega_{0}$. With this choice and in the \emph{scaling limit} $\hbar\omega_{c}\gg k_{B}T$, the function $Q(t)$ acquires the form~\cite{Weiss2012}
\begin{equation}\label{Q_t}
\begin{aligned}
Q(t)=S(t)+iR(t)=&\frac{M\gamma}{\pi\hbar}\ln\left(\sqrt{1+\omega_{c}^{2} t^{2}}\frac{\sinh(\kappa t)}{\kappa t}\right)\\
&+i\frac{M\gamma}{\pi\hbar}\arctan(\omega_{c}t),
\end{aligned}
\end{equation}
where $\kappa=\pi k_{B}T/\hbar$.\\
\indent In this work we calculate the populations of the localized states $|Q_{k}\rangle$, assuming that the particle is initially in the state $|Q_{1}\rangle$ (see Fig.~\ref{fig1}). This amounts to considering $q=q'=Q_{k}$ and $\rho_{q_{0}q_{0}'}(t_{0})=\delta_{q_{0}'q_{0}}\delta_{q_{0}Q_{1}}$in Eq.~(\ref{RDM-DVR}), which reduces to
\begin{equation}\label{rho_t}
\begin{aligned}
\rho_{kk}(t)&\equiv\langle Q_{k}|\rho|Q_{k}\rangle \\
&= G(Q_{k},Q_{k},t;Q_{1},Q_{1},t_{0}),\qquad k=1,\dots,4.
\end{aligned}
\end{equation}
Two major difficulties arise in the evaluation of the propagator: the variety of possible paths of the multi-state system and  the intricate time nonlocal correlations among the $\xi$- and $\chi$-charges introduced by $\mathcal{F}_{FV}$.\\
\indent In the next section a novel approximation scheme, which overcomes these difficulties  for the intermediate to high temperature and intermediate to strong coupling, is introduced.
\section{The VR-WIBA approximation}
\label{approximations}
In this section we introduce a novel approximation scheme, valid in the crossover regime of damped intrawell oscillations and incoherent tunneling. The derivation is in three stages. First, we retain only the leading contributions to the sum over paths in Eq.~(\ref{propagator}). Second, a class of interactions in the influence functional for the leading paths is neglected. This is done on the basis of the difference in time scales characterizing the system. Finally, the time nonlocal part of the interactions retained is treated to the first order in the coupling. The resulting approximation scheme is introduced in Sec.~\ref{VR-WIBA}: It combines a NIBA-like approach for the tunneling dynamics and a more refined scheme for the intrawell motion,  capturing the non-Markovian character of the reduced dynamics.  At high temperatures this scheme reduces to the preexisting  NIBA-like approaches for multi-state systems~\cite{ThorwartPRL2000}.
\subsection{Selection on the paths}
\label{path-selection}
\begin{figure}[htbp]
\begin{center}
\includegraphics[width=4.5cm,angle=0]{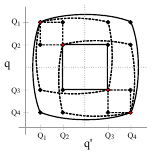}
\caption{\small{(Color online) Within the approximations made in Sec.~\ref{path-selection}, the motion of the system in the $4\times4$ grid of spatial positions decouples into a sequence of arbitrarily long  paths along the six two-dimensional square sublattices shown in the figure. Each square  represents a different two-level system characterized by its own bias,  tunneling transition amplitude, and spatial distance between the states. Dashed and solid sublattices indicate biased and unbiased effective two-level systems, respectively. For example, paths in the external sublattice represent tunneling transitions between the states $|Q_{1}\rangle$ and $|Q_{4}\rangle$.}}
\label{fig6}
\end{center}
\end{figure}

A diagonal path configuration ($q=q'$) is called \emph{sojourn}. The non-diagonal configurations ($q\neq q'$) between two consecutive sojourns for a multi-state system constitute a so-called \emph{cluster}, which is a generalization of the \emph{blip}, a single off-diagonal excursion of the reduced density matrix~\cite{Leggett1987}. In the charge picture mentioned above, clusters are neutral objects because, inside a cluster, the charges sum up to zero. In Fig.~\ref{fig5} we give an example of a path with three sojourns and two clusters, the first of which is a simple blip while the second has multiple off-diagonal transitions.\\ 
\indent The leading contributions to the sum over paths in Eq.~(\ref{propagator}) are from paths returning in a sojourn configuration after a \emph{single} off-diagonal excursion (blip). This is because long clusters are suppressed by the real part of the influence phase, which produces a cutoff through $\xi_{j}S_{jj-1}\xi_{j-1}=-(q_{j}-q'_{j})^{2}S(t_{j}-t_{j-1})$ (see Eq.~(\ref{FV-DVR})). The cutoff depends on the tunneling distance $|q_{j}-q'_{j}|$ and on the coupling strength $\gamma$, which enters as a pre-factor in $S(t)$  (see Eq.~(\ref{Q_t})).\\
\indent  The first approximation we make is to retain exclusively the above mentioned leading contributions, i.e., the paths with the simplest possible clusters, namely \emph{vibrational relaxation}-blips (VR-blips) and \emph{tunneling}-blips (T-blips).  In a VR-blip the off diagonal configuration is with $q$ and $q'$ belonging to the same well, whereas in a T-blip $q$ and $q'$ belong to different wells. The validity of this approximation rests on the effectiveness of the $\gamma$-dependent cutoff in suppressing the long clusters and is therefore guaranteed at sufficiently strong coupling.\\ 
\indent Under the approximation discussed above, a generic path is thus a sequence of alternating blips and sojourns along the six sublattices depicted in Fig.~\ref{fig6}. Each sublattice corresponds to the coordinate space of a two-level system characterized by its own bias $\epsilon$, transition amplitude $\Delta$ per unit time  (see Eqs.~(\ref{delta-j}) and~(\ref{epsilon-j})), and spatial distance between the states.\\
\subsection{Retained interactions}
\label{interactions}
\begin{figure}[htbp]
\begin{center}
\includegraphics[width=8cm,angle=0]{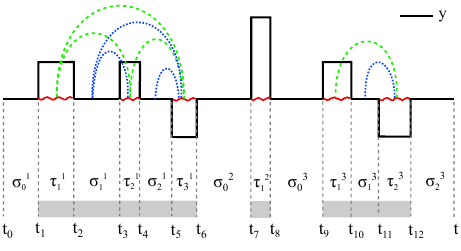}
\caption{\small{(Color online) A path of the coordinate $y=q-q'$ made by two VR-blip sequences (of lengths $t_{6}-t_{1}$ and $t_{12}-t_{9}$) separated by a T-blip of length $t_{8}-t_{7}$. The transition times and the blip/sojourn times are indicated. The shaded areas on the lower part of the figure represent the time intervals inside which the correlations are retained according to the approximations discussed in Sec. \ref{path-selection}. Specifically, according to the VR-WIBA scheme introduced in Sec. \ref{VR-WIBA}, the intra-VR-blip and intra-T-blip interactions (solid wavy lines) are taken at all orders in $\gamma$, while the inter-VR-blip and VR-blip-sojourn interactions (dashed lines) are taken to the first order in $\gamma$.}}
\label{fig7}
\end{center}
\end{figure}

\indent The second approximation is on the blip-blip and blip-sojourn interactions in the influence phase (Eq.~(\ref{FV-DVR})) for the retained paths. Specifically, we neglect the interactions between
\begin{enumerate}
\item T-blips,
\item T-blips and VR-blips, 
\item VR-blips separated by T-blips.
\end{enumerate} 
This approximation is justified by the long time scale of the tunneling dynamics, i.e., by the fact that, on average, tunneling transitions are rare, because their amplitudes per unit time are small compared to those of  the vibrational relaxation (see Eqs.~(\ref{Delta-DDS}) and~(\ref{Delta-DDS2}) in Appendix~\ref{parameters}).
Since in a typical path the T-blips are well separated in time, they interact through the following long time (or high temperature) limit of Eq.~(\ref{Q_t})~\cite{Thorwart2001} 
\begin{equation}\label{Q-long-time}
Q(t)=S(t)+iR(t) =\frac{M\gamma}{\pi\hbar}\left[\kappa t -\ln\left(\frac{2\kappa}{\omega_{c}} \right)\right]+i\frac{M\gamma}{2\hbar}.
\end{equation}
In this limit the T-blips decouple exactly, due to the linearity of $Q(t)$~\cite{Thorwart2001}.  The difference in magnitude of the spatial distances between the localized states involved in the transitions produces a different effective damping for the intra- and interwell dynamics. As a result tunneling blips are strongly suppressed by the large $(q_{j}-q'_{j})^{2}$ factor present in the real part of the influence phase.\\
\indent The picture of a typical path of the class retained is that of a sequence of interacting VR-blips, interrupted by isolated T-blips, as sketched in Fig.~\ref{fig7}. Notice that, inside a sequence of consecutive VR-blip, the interactions are fully retained.\\      
\subsection{Weakly interacting VR-blip approximation}
\label{VR-WIBA}
The last approximation regards the intricate time non-local interactions among VR-blips.\\
\indent At temperatures $T\sim\hbar\Omega_{0}/k_{B}$ the linear form~(\ref{Q-long-time}) for $Q(t)$ is attained also on the characteristic time scale of  VR-blips ($\sim\Omega_{0}^{-1}$, see  Eq.~(\ref{Delta-DDS})). As a result, VR-blips decouple and it is a good approximation to retain exclusively the intra-blip interactions represented by the wavy red lines in Fig.~\ref{fig7}. The resulting overall scheme is the multi-level generalization of the NIBA, which can be called \emph{generalized non-interacting blip approximation} (\textbf{gNIBA}). This scheme coincides with the generalized non-interacting cluster approximation~\cite{Thorwart2001} at the leading order in the coefficients $\Delta_{ij}$.\\ 
\indent  If the temperature is not sufficiently high to ensure the decoupling of the VR-blips, the gNIBA fails and inter-blip interactions within a VR-blips sequence (dashed lines in Fig.~\ref{fig7}) must be considered.\\ 
\indent  In this case the time nonlocal interactions in the vibrational relaxation dynamics (intrawell motion) can be accounted for perturbatively by using the \emph{weakly interacting blip approximation} (WIBA)~\cite{NesiPRB2007}. The WIBA interpolates between the strong and weak coupling regimes by treating to the first order in $\gamma$ the time nonlocal inter-blip interactions and at all orders the intra-blip interactions. \\
\indent In the resulting overall scheme the Feynman-Vernon influence functional is approximated according to the NIBA for the tunneling transitions and to the WIBA for the intrawell transitions. 
For this reason we call the scheme \emph{weakly interacting VR-blip approximation} (VR-WIBA)~\cite{NesiPhD2007}.\\ 
\indent Two considerations are in order about this novel  approximation scheme. The first is that its domain of validity in the parameter space includes that of the gNIBA (see Fig.~\ref{fig9}). This is because the complete decoupling of the blips in the gNIBA implies the separation among T-blips and VR-blips, which is at the basis of the VR-WIBA. Second, in establishing the validity of  both schemes not only the temperature, but also the coupling strength plays an important role. This is because neglecting contributions from long clusters in the sum over paths is possible as long as they are suppressed by a sufficiently large coupling with the environment, as discussed in Sec.~\ref{path-selection}. Moreover, the treatment of the inter-VR-blip interactions to the first order in the coupling makes the VR-WIBA unsuited for the low-temperature-and-strong-coupling regime. Further discussions about the validity domains of approximation schemes are deferred to Sec.~\ref{phase diagram}, where the phase diagram in the coupling-temperature space is presented.
\section{VR-WIBA generalized master equation}
\label{GME-vrwiba}
According to the approximations made in Sec.~\ref{path-selection}, a path of the reduced density matrix is a sequence of \emph{non-interacting}  subpaths of six different two-level systems (TLSs) corresponding to the $2\times2$ sublattices depicted in Fig.~\ref{fig6}.\\ 
\indent We are interested in calculating the populations $\rho_{kk}$ ($k=1,\dots,4$) with initial condition $\rho(t_{0})=|Q_{1}\rangle\langle Q_{1}|$. As a consequence, the initial and final states are the diagonal states $q_{0}=q_{0}'=Q_{1}$ and $q=q'=Q_{k}$ (see Eq.~(\ref{rho_t})). It follows that each subpath has an even number $2k$ of transitions, and that the total number of transitions in a path is the even number $2n$.\\
\indent  Consider a path with $2n$ transitions distributed in $N$ subpaths, each of which has $2k_{j}$ transitions ($j=1,\dots,N$). For what stated above the amplitude $\mathcal{A}[q]\mathcal{A}^{*}[q']\mathcal{F}_{FV}[x,y]:=A(\tau,\sigma)$ of this path factorizes as  
\begin{equation}\label{factorization-amplitude}
\begin{aligned}
A(\tau,\sigma)=\prod_{j=1}^{N}A_{j}(\tau_{1}^{j},\sigma_{1}^{j},\dots,\sigma_{k_{j}-1}^{j},\tau_{k_{j}}^{j}),
\end{aligned}
\end{equation}
where $\tau_i=t_{2i}-t_{2i-1}$ are the blip times and $\sigma_i=t_{2i+1}-t_{2i}$ the sojourn times depicted in Fig.~\ref{fig7}.\\ 
\indent  The population  $\rho_{kk}(t)$ of the state $|Q_{k}\rangle$ at time $t$, with initial condition $\rho(t_{0})=|Q_{1}\rangle\langle Q_{1}|$, is (see Eq.~(\ref{propagator}))
\begin{equation}\label{rho_t_DDS}
\begin{aligned}
\rho_{kk}(t)=&\delta_{k,1}+\sum_{n=1}^{\infty}\int_{t_0}^{t}\mathcal{D}_{2n}\{\tau,\sigma\}\prod_{j=1}^{N}A_{j}(\tau_{1}^{j},\sigma_{1}^{j},\dots,\tau_{k_{j}}^{j}),
%A(t_{0},t_{1},\dots,t_{2n},t).
\end{aligned}
\end{equation}
where $\sum_{j=1}^{N}2k_{j}=2n$ and where
\begin{equation}\label{int-D-b}
\begin{aligned}
\int_{t_{0}}^{t}&D_{2n}\{\tau,\sigma\}=\sum_{\text{paths}_{2n}}\int_0^{t-t_{0}} d\sigma_0\int_0^{t-t_{0}-\sigma_0} d\tau_1\dots\\
&\times\int_0^{t-t_{0}-\dots-\tau_{n-1}} d\sigma_{n-1}\int_0^{t-t_{0}-\dots-\sigma_{n-1}} d\tau_{n}. 
\end{aligned}
\end{equation}
\indent Let the $j$-th subpath start in the diagonal site ($q_{j},q_{j}$) and end  in the diagonal site ($q_{j+1},q_{j+1}$): Due to the factorization in Eq.~(\ref{factorization-amplitude}) the Laplace  transform of $\rho_{kk}(t)$ reads
\begin{equation}\label{rho_lambda_DDS}
\begin{aligned}
\hat{\rho}_{kk}(\lambda)=&\frac{\delta_{k,1}}{\lambda}+\frac{1}{\lambda}\sum_{N=1}^{\infty}\sum_{\{q\}=Q_{1}}^{Q_{4}}\hat{g}_{q_{1},q_{2}}(\lambda)\\
&\times\hat{g}_{q_{2},q_{3}}(\lambda)\dots\hat{g}_{q_{N},Q_{k}}(\lambda)\delta_{q_{1}Q_{1}},
\end{aligned}
\end{equation}
where the sum is over the set $\{q\}=q_{1},\dots,q_{N}$.
The $1/\lambda$ factors in Eq.~(\ref{rho_lambda_DDS}) appear after integration over sojourn times, as described in Appendix~\ref{propagator-laplace}.
The function $\hat{g}_{q_{j},q_{j+1}}$ is related to the Laplace transform of  the two-level system propagator from ($q_{j},q_{j}$) to ($q_{j+1},q_{j+1}$)
\begin{equation}\label{prop-TLS}
\begin{aligned}
G_{q_{j},q_{j+1}}(t)=\sum_{k_{j}=1}^{\infty}\int_{0}^{t}\mathcal{D}_{2k_{j}}\{\tau,\sigma\}A_{j}(\tau_{1}^{j},\sigma_{1}^{j},\dots,\tau_{k_{j}}^{j}).
\end{aligned}
\end{equation}
If $q_{j+1}\neq q_{j}$, then the two-level system is identified by  the values  of $q_{j}$ and $q_{j+1}$. The following relation holds (see Appendix~\ref{propagator-laplace})
 \begin{equation}\label{propagatorTLS-laplace}
\begin{aligned}
\hat{g}_{q_{j},q_{j+1}}&(\lambda)=\lambda\hat{G}_{q_{j},q_{j+1}}(\lambda)\\
=&\delta_{s,r}+\sum_{n=1}^{\infty}\sum_{\{\nu\}=r,s}\frac{\hat{K}_{\nu_{1},\nu_{2}}(\lambda)}{\lambda}
\dots\frac{\hat{K}_{\nu_{n},s}(\lambda)}{\lambda}\delta_{\nu_{1},r},
 \end{aligned}
\end{equation}
where $\hat{K}$ are irreducible two-level system kernels (see Appendix~\ref{TLSpropagator}).\\
\indent It follows that the Laplace transform of the population $\rho_{kk}$, given in Eq.~(\ref{rho_lambda_DDS}), can be recast in the form
 \begin{equation}\label{rho_lambda_DDS2}
\begin{aligned}
\hat{\rho}_{kk}(\lambda)=&\frac{\delta_{k,1}}{\lambda}+\frac{1}{\lambda}\sum_{n=1}^{\infty}\sum_{\{q\}=Q_{1}}^{Q_{4}}\frac{\hat{K}_{q_{1},q_{2}}(\lambda)}{\lambda}\\
&\times\frac{\hat{K}_{q_{2},q_{3}}(\lambda)}{\lambda}\dots\frac{\hat{K}_{q_{n},Q_{k}}(\lambda)}{\lambda}\delta_{q_{1}Q_{1}}.
 \end{aligned}
\end{equation}
\indent We now switch to vector notation by defining the four-dimensional vector $\vec{\rho}(\lambda)$, whose components are $\hat{\rho}_{kk}(\lambda)$, and  the $4\times4$ matrix $\hat{\mathcal{K}}(\lambda)$, whose off-diagonal elements are the TLS kernels $\hat{K}(\lambda)$. Within this notation  Eq.~(\ref{rho_lambda_DDS2}) reads
\begin{equation}
\begin{aligned}\label{gme-laplace}
\vec{\rho}(\lambda)=&\frac{\vec{\rho}(t_{0})}{\lambda}+\frac{1}{\lambda}\sum_{N=1}^{\infty}\left[\frac{\mathcal{\hat{K}}(\lambda)}{\lambda} \right]^N\vec{\rho}(t_{0})\\
=&\frac{1}{\lambda}\sum_{N=0}^{\infty}\left[\frac{\mathcal{\hat{K}}(\lambda)}{\lambda} \right]^N\vec{\rho}(t_{0})\\
=&\left[\lambda\mathbb{I}-\mathcal{\hat{K}}(\lambda)\right]^{-1}\vec{\rho}(t_{0}).
\end{aligned}
\end{equation}
Transforming back to the time domain we obtain the following generalized master equation (GME) 
\begin{equation}\label{vec-rho}
\dot{\vec{\rho}}(t)=\int_{t_0}^{t}dt' \mathcal{K}(t-t')\vec{\rho}(t'),
\end{equation}
where $\mathcal{K}(t)=\mathcal{L}^{-1}\{\mathcal{\hat{K}}(\lambda)\}$. Restoring the index notation in Eq.~(\ref{vec-rho}) we have
 \begin{equation}\label{GME}
 \dot{\rho}_{kk}(t)=\sum_{j=1}^{4}\int_{t_0}^{t}dt' K_{kj}(t-t')\rho_{jj}(t'),
 \end{equation} 
where $K_{kj}\equiv K_{Q_{k}Q_{j}}$. The diagonal elements of the kernel matrix $\mathcal{K}(t)$ are given by probability conservation 
 \begin{equation}
K_{jj}(t)=-\sum_{\substack{
   j=1 \\
   j\neq i}
  }^{4} K_{i j}(t).
\end{equation} 
\indent  In the VR-WIBA scheme the kernels $K_{kj}$ are hence as follows:
\begin{itemize}
 \item If  the states $|Q_{k}\rangle$ and $|Q_{j}\rangle$ connected by $K_{kj}$ belong to different wells then the corresponding kernel is a NIBA kernel $K_{kj}^{N}(t)$. 
\item If $|Q_{k}\rangle$ and $|Q_{j}\rangle$ belong to the same well then a WIBA kernel, which consists of a NIBA plus a beyond-NIBA part 
$K_{kj}^{W}(t)=K_{kj}^{N}(t)+K_{kj}^{BN}(t)$,  is used.
Setting to zero the beyond-NIBA correction in these intrawell kernels amounts to use the gNIBA~\cite{Grifoni1996,Thorwart2001} scheme for the full system. 
 \end{itemize}
The explicit expressions for the kernels are given in Appendix~\ref{vrwiba-kernels}.\\ 
\indent  At strong coupling and high temperature, in the incoherent regime  where the gNIBA is appropriate (see Sec.~\ref{phase diagram}), populations evolve on time scales larger than the time intervals over which the gNIBA kernels substantially differ from zero.  This observation suggests that we cast Eq.~(\ref{GME})  into the Markov approximated master equation form
\begin{equation} 
\label{GME-markov}
 \dot{\rho}_{kk}(t)=\sum_{j=1}^{4}\Gamma_{kj}\rho_{jj}(t), \quad \text{where} \quad \Gamma_{kj}=\int_{0}^{\infty}dt K_{kj}^{N}(t).
 \end{equation}
Equation~(\ref{GME-markov}) describes well the incoherent regime occurring at strong coupling. The analytical solution is of the form $\rho_{kk}(t)=\sum_{i,j}a_{ki}b_{ij}\exp[\Lambda_{i}(t-t_{0})]\rho_{jj}(t_{0})$.\\
\indent Let $\Lambda_{\text{min}}$ be the smallest, in absolute value, of the nonzero eigenvalues of the rate matrix $\Gamma$. The rate $|\Lambda_{\text{min}}|$ the so-called \emph{quantum relaxation rate}~\cite{Thorwart2001}. The relaxation time
\begin{equation}\label{relax-time}
\tau_{R}=|\Lambda_{\text{min}}|^{-1}
\end{equation}
sets the time scale of the relaxation towards equilibrium.\\
\begin{figure}[htbp]
\begin{center}
\includegraphics[width=8cm ,angle=0]{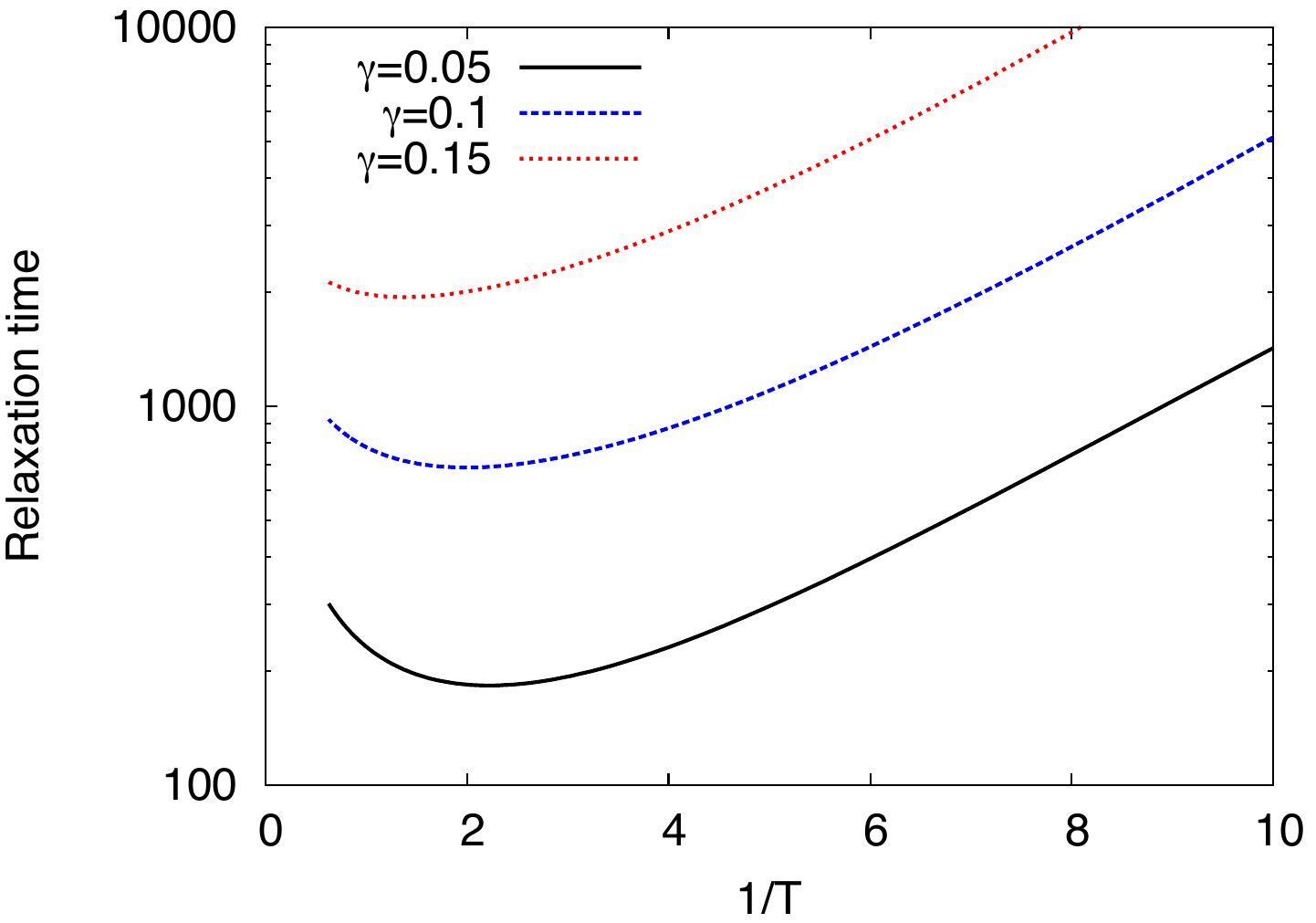}
\caption{\small{(Color online) Relaxation time (given by Eq.~(\ref{relax-time})) in units of $\omega_{0}^{-1}$, vs inverse temperature for different values of the coupling strength $\gamma$. Temperature and $\gamma$ are in units of $\hbar\omega_{0}/k_{B}$ and $\omega_{0}$, respectively.}}
\label{fig8}
\end{center}
\end{figure}
\indent As an application of our model, we calculate the relaxation time, given by Eq.~(\ref{relax-time}), as a function of the inverse temperature for three values of  $\gamma$. The potential used is that given in Fig.~\ref{fig2} and the dissipation is Ohmic with $\omega_{c}=50~\omega_{0}$ (see Eq.~(\ref{J-continuous})). The results, shown in Fig.~\ref{fig8}, display a minimum in the relaxation time reproducing an activated rate, which can be ascribed to the presence of the higher energy doublet, and is qualitatively similar to that found in Ref.~\cite{Cukier1990} in the context of proton transfer reactions.\\
\indent In the remaining of the paper we shall investigate various dynamical regimes associated to the phase diagram introduced in the next section.
\section{Dynamical regimes, approximation schemes, and physical realizations}
\label{phase diagram}
In what follows we use different techniques, notably the VR-WIBA generalized master equation derived in the preceding section, to obtain the dynamics of the double-doublet system (DDS) with Ohmic dissipation. But before doing this it is useful to give a general overview of the various dynamical regimes corresponding to different regions of the parameter space of the DDS. To this end we present here a phase diagram in the ($\gamma,T$)-space, which summarize the dynamical regimes along with the domains of validity of the techniques used.\\
\subsection{Phase diagram}
\label{PD}
 As a starting point we note that the dynamics of a TLS occurs in the coherent or incoherent tunneling regime~\cite{Weiss2012}, while for multi-level systems there is a richer variety of dynamical behaviors due to the different energy scales involved. The presence of multiple energy scales in our problem is best seen in the free dynamics depicted in Fig.~\ref{fig3}, which displays the three characteristic frequencies $\Omega_2\ll\Omega_1<\Omega_0$. As discussed in Sec. \ref{DDS}, this corresponds to a slow and a fast tunneling dynamics, occurring on the time scales $\Omega_2^{-1}$ and $\Omega_1^{-1}$, respectively,  and to intrawell oscillations of frequency  $\Omega_0$.\\
\indent In the presence of a dissipative environment, these different oscillatory behaviors undergo coupling and temperature-dependent frequency shifts, and are progressively suppressed at different dissipation regimes.  \\ 
\indent  We exploit the decoupling of the DDS into effective TLSs,  in conjunction with numerical tests, to identify the following dynamical regimes in the $(\gamma,T)$-space:
\begin{itemize}
\item[] $A$ - completely coherent regime, with coherent tunneling oscillations at both the long ($\Omega_2^{-1}$) and the intermediate ($\Omega_1^{-1}$) time scales, and coherent intrawell oscillations on the short time scale $\Omega_0^{-1}$;
 \item[] $B$ -  coherent tunneling on the time scale $\Omega_1^{-1}$ set by the higher energy doublet and coherent intrawell dynamics. Oscillations of the left- or right-well populations around a slow incoherent relaxation behavior  associated to the lower doublet; 
\item[] $C$ - crossover regime, where the coherence is only at the level of  intrawell motion (on the time scale $\Omega_{0}^{-1}$) and  tunneling is incoherent;
\item[]  $D$ incoherent regime, where the four populations relax incoherently to their equilibrium values.
\end{itemize}
\subsection{Approximation schemes}
\label{approx-schemes}
\begin{figure}%[htbp]
\begin{center}
\includegraphics[width=8.7cm ,angle=0]{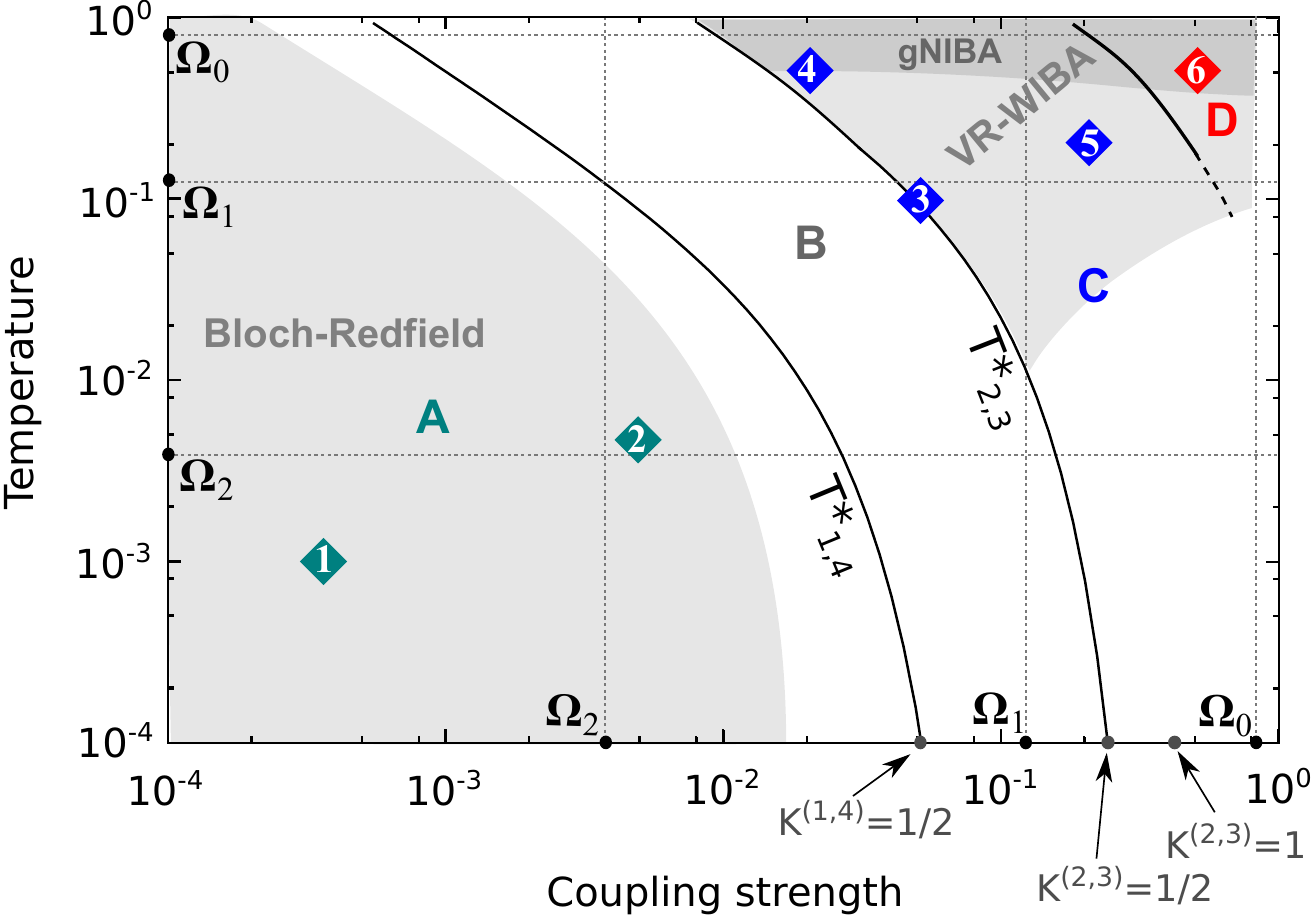}
\caption{\small{(Color online) Phase diagram of the dissipative double-doublet system in the coupling-temperature plane. Dynamical regimes $A$-$D$ are separated by solid lines. Region $B$ is delimited by the curves $T^{*}_{1,4}(\gamma)$ and $T^{*}_{2,3}(\gamma)$, the coherent-incoherent transition temperatures for the two-level systems $\{|Q_1\rangle,|Q_4\rangle\}$ and $\{|Q_2\rangle,|Q_3\rangle\}$, respectively (see Eqs.~(\ref{kondo}) and~(\ref{T-star})).
Shaded areas indicate the validity domains of Bloch-Redfield and of path-integral approaches within the VR-WIBA and gNIBA. The characteristic frequencies $\Omega_{i}$, in units of $\omega_{0}$, are shown as a reference. Diamonds denote the six phase points chosen for the results presented in Sec.~\ref{results}. 
Values of $\gamma$ for which the effective couplings $K^{(1,4)}$ and $K^{(2,3)}$ assume specific values are indicated.   
Temperatures and couplings are in units of $\hbar\omega_{0}/k_{B}$ and $\omega_{0}$, respectively.}}
\label{fig9}
\end{center}
\end{figure}

\indent  The corresponding regions in the parameter space along with the validity areas of the approximation schemes used are depicted in the phase diagram of Fig.~\ref{fig9}.\\
\indent At intermediate coupling and temperature the dynamical regimes are controlled by the behavior of the effective TLSs introduced in Sec.~\ref{approximations}, as the natural description for the system passes from the energy representation to the localized one given by the discrete variable representation. It is thus suggestive to use  the machinery existing for the spin-boson problem to give an indication of the boundaries in the parameter space between different dynamical regimes.\\
%x
\indent  As discussed in Sec.~\ref{approximations}, each of the six effective TLSs, which we denote by $\{|Q_i\rangle,|Q_j\rangle\}$ ($i\neq j$),  has its tunneling element $\Delta_{ij}$, bias $\epsilon_{ij}$ (see Appendix~\ref{parameters} for numerical values), and characteristic distance $q_{ij}=Q_{i}-Q_{j}$. As a consequence, at fixed $\gamma$, the effective coupling strength for an effective TLS can be more or less strong, depending on the TLS considered. 
This effective coupling strength, also called  Kondo parameter, for $\{|Q_i\rangle,|Q_j\rangle\}$ is defined by 
\begin{equation}\label{kondo}
K^{(i,j)}=M\gamma q_{ij}^{2}/(2\pi \hbar).
\end{equation}
At $T=0$, for a \emph{symmetric} two-level system, $K=1/2$ corresponds  to the transition from the coherent to the incoherent behavior, while at $K=1$ \emph{localization} occurs, consisting in the complete inhibition of tunneling, so the particle does not leave the well where it was prepared. \\
\indent The coherent-incoherent transition temperature $T^{*}$ as a function of $K$ for a symmetric TLS is given by
\begin{equation}\label{T-star}
T^{*}(K)=\left(\frac{(2\pi)^{K}}{\pi K}\right)^{1/(1-K)}\frac{\hbar \Delta_{r}}{k_{B}},
\end{equation}
where $\Delta_{r}=\Delta(\Delta/\omega_{c})^{K/(1-K)}$ ($K<1$) is the renormalized value of the tunneling element $\Delta$~\cite{Weiss2012}.\\ 
\indent In our system the two effective TLSs $\{|Q_1\rangle,|Q_4\rangle\}$ and $\{|Q_2\rangle,|Q_3\rangle\}$ are both characterized by $\epsilon_{ij}=0$ and are thus symmetric (see Eq.~(\ref{epsilon-DDS})).
 The values of $\gamma$ corresponding to $K=1/2$ for these two TLSs, namely $\gamma(K^{(1,4)}=1/2)$ and  $\gamma(K^{(2,3)}=1/2)$, are shown in Fig.~\ref{fig9} along with $\gamma(K^{(2,3)}=1)$ for $\{|Q_2\rangle,|Q_3\rangle\}$.\\
\indent  In the phase diagram of Fig.~\ref{fig9} the boundary of region $A$ of complete coherence is established by considering $T^{*}_{1,4}$, the  coherent-incoherent transition temperature as a function of $\gamma$ for the effective TLS  $\{|Q_1\rangle,|Q_4\rangle\}$ (see Eqs.~(\ref{kondo}) and~(\ref{T-star})). Indeed, $T^{*}_{1,4}$ determines the suppression of the  slow tunneling oscillations of frequency $\Omega_{2}$, while both tunneling oscillations of frequency $\Omega_{1}$ and the fast intrawell dynamics survive.\\ 
\indent  In the part of region $A$ where the perturbative in the coupling treatment is appropriate, the Bloch-Redfield master equation describes correctly the time evolution of the reduced density matrix. The solutions for the coherences in the energy representation (see Appendix~\ref{bloch-redfield}) are
\begin{equation}\label{solution-coherences}
 \rho_{nm}^{E}(t)=e^{-i\omega_{nm}t}e^{-\mathcal{L}_{nm,nm}t}\rho_{nm}^{E}(t_{0}).
\end{equation}
To determine the domain of validity  of the Bloch-Redfield approach we compare the dephasing coefficient $\mathcal{L}_{12,12}$ with $\omega_{21}\equiv\Omega_{2}=\omega_{2}-\omega_{1}$ and require that $\mathcal{L}_{12,12}\leq\omega_{21}$. The result is the shaded area on the left part of the phase diagram. However, the extent of this validity domain may be 
overestimated, having neglected the frequency shifts in the evaluation of the Bloch-Redfield tensor (see. Eq.~(\ref{principal-value})).  
The weak-coupling approximation fails near the boundary between $A$ and $B$ where the low-frequency oscillations turn into incoherent relaxation. \\ 
\indent  Region $B$ of the diagram, characterized by stronger coupling and/or higher temperature, is outside the validity domain of the Bloch-Redfield approach. Nevertheless, we can use the weak coupling estimates for the asymptotic values of the populations on the basis of the following argument. Due to the symmetry of the system, the asymptotic values of the left- and right-well population are $P_{L}(\infty)=P_{R}(\infty)=1/2$, where $P_{L( R )}=\rho_{11(33)}+\rho_{22(44)}$. As a consequence the individual populations $\rho_{ii}(\infty)$ at equilibrium can be given focusing on the intrawell effective TLSs $\{|Q_1\rangle,|Q_2\rangle\}$ and $\{|Q_3\rangle,|Q_4\rangle\}$. These two-level systems are characterized by effective tunneling element $|\Delta_{12}|=|\Delta_{34}|\simeq 0.35~\omega_0$, effective bias $|\epsilon_{12}|=|\epsilon_{34}|\simeq 0.40~\omega_0$, and small distance between the states $|q_{12}|=|q_{34}|\simeq1.69~\sqrt{\hbar/(M\omega_{0})}$. Therefore, because of Eq.~(\ref{kondo}), the effective couplings $K^{(1,2)}=K^{(3,4)}$ are weak and we can use the  following weak coupling expressions~\cite{Weiss2012} for the asymptotic populations 
\begin{equation}\label{rho_asympt}
\begin{aligned}
\rho_{33/44}(\infty)=&\rho_{22/11}(\infty)\\
=&\frac{1}{4}\mp\frac{\epsilon}{4\Delta_{b}}\tanh\left(\frac{\hbar\Delta_{b}}{2k_{B}T}\right),\\
\end{aligned}
\end{equation}
where $\Delta_{b}=\sqrt{\Delta_{12}^{2}+\epsilon_{12}^{2}}$. The second line of Eq.~(\ref{rho_asympt}) derives from the symmetry of the problem. In Sec.~\ref{results} we compare the asymptotic values of the populations obtained by the VR-WIBA generalized master equation in the crossover regime $C$ with those given by Eq.~(\ref{rho_asympt}).\\ 
\indent The boundary between regions $B$ and $C$ is the transition to the crossover regime in which also the tunneling associated to the higher energy doublet is incoherent. The  boundary is obtained by considering the coherent-incoherent transition temperature $T^{*}_{2,3}$  as a function of $\gamma$ for the symmetric TLS $\{|Q_{2}\rangle,|Q_{3}\rangle\}$.\\ 
\indent The crossover region $C$ and the fully incoherent region $D$ define also the range in which the contributions from long clusters in the sum over paths are negligible.\\ 
\indent  Since the two intrawell effective two-level systems have an effective bias, they can be treated according to the NIBA only in the high-temperature/strong-coupling regime (on the frequency scale $\Omega_{0}$ of the intrawell motion)~\cite{Weiss2012}.
 Treating the intrawell dynamics within the NIBA amounts to apply for the complete system the gNIBA, which reproduces correctly the dynamics inside the darker shaded area in the uppermost part of the diagram.\\ 
\indent  The VR-WIBA extends the path-integral approach for the double-doublet system to low temperatures in a quite large range of coupling strengths. The validity area of the VR-WIBA includes that of the gNIBA, as discussed in Sec.~\ref{VR-WIBA}, and covers the upper-right shaded region in the phase diagram.\\ 
\indent  Dissipation regimes not accessible to the VR-WIBA in regions $C$ and $D$ are in the low temperature and strong coupling regime, corresponding to the lower-right part of the phase diagram. There, the inter-blip correlations are not suppressed by the bath and the coupling is not sufficiently weak to justify their treatment to the first order in $\gamma$.
\subsection{Physical realizations}
\label{experiments}
Real physical systems modeled as multi-level bistable systems are found in several areas. Examples are molecular nanomagnets, high-spin molecules displaying  tunneling of the magnetization between two potential wells separated by a large barrier~\cite{Gatteschi2006}. These molecules feature a number of energy levels under the barrier which corresponds to the projections of the total spin along a preferred direction. Incoherent tunneling of magnetization has been experimentally observed in the Mn$_{12}$ molecule~\cite{Friedman1996}.\\
\indent  The specific model considered in this work, with a couple of energy doublets under the barrier, effectively describes experiments on proton transfer reactions in benzoic acid crystals. Indeed, the experimental curves in Ref.~\cite{Nagaoka1983} on the proton relaxation time as a function of temperature are theoretically reproduced in Ref.~\cite{Cukier1990} by using a double-doublet system linearly coupled to a harmonic bath.\\ 
\indent Archetypal systems, whose theoretical description is based on the model investigated here, are superconducting quantum interference devices (SQUIDs),    superconducting rings interrupted in one or more points by thin layers of insulator. In these devices, of interest in quantum computation, the magnetic flux associated to the current threading the circuit is subject to an effective bistable potential with quantized energy levels. Incoherent tunneling of the magnetic flux through the potential barrier  at strong dissipation has been first observed in SQUIDs in  Ref.~\cite{Han1991}, and coherent tunneling dynamics has been demonstrated in Refs.~\cite{Chiorescu2003,Friedman2000}.\\
\indent SQUID-based flux qubits, operated by external fluxes, allow for the manipulation of both the bias and barrier height of the double well potential. In Ref.~\cite{Poletto2009} these parameters are tuned by using fast dc pulses and a protocol is realized for measuring oscillations of the left- or right-well populations. The protocol consists in preparing the state of the qubit in a well by applying a large bias and then restoring a symmetric configuration with a large potential barrier to prevent the tunneling to the other well on the protocol's time scale. The barrier is then removed and further restored to perform a \emph{which well} measurement. The flexibility of the device makes it possible to attain the double-doublet configuration studied here by imposing a suitable barrier height and also to probe the dynamics with nonequilibrium initial conditions.
\section{Dissipative dynamics of the double-doublet system}
\label{results}
\subsection{Parameters and units}
\label{preamble-results} 
Throughout this section parameters are scaled with $\omega_{0}$, the oscillation frequency around the minima of the potential described by Eq.~(\ref{Potential}).
\begin{figure}[htbp]
\begin{center}
\includegraphics[width=8.5cm ,angle=0]{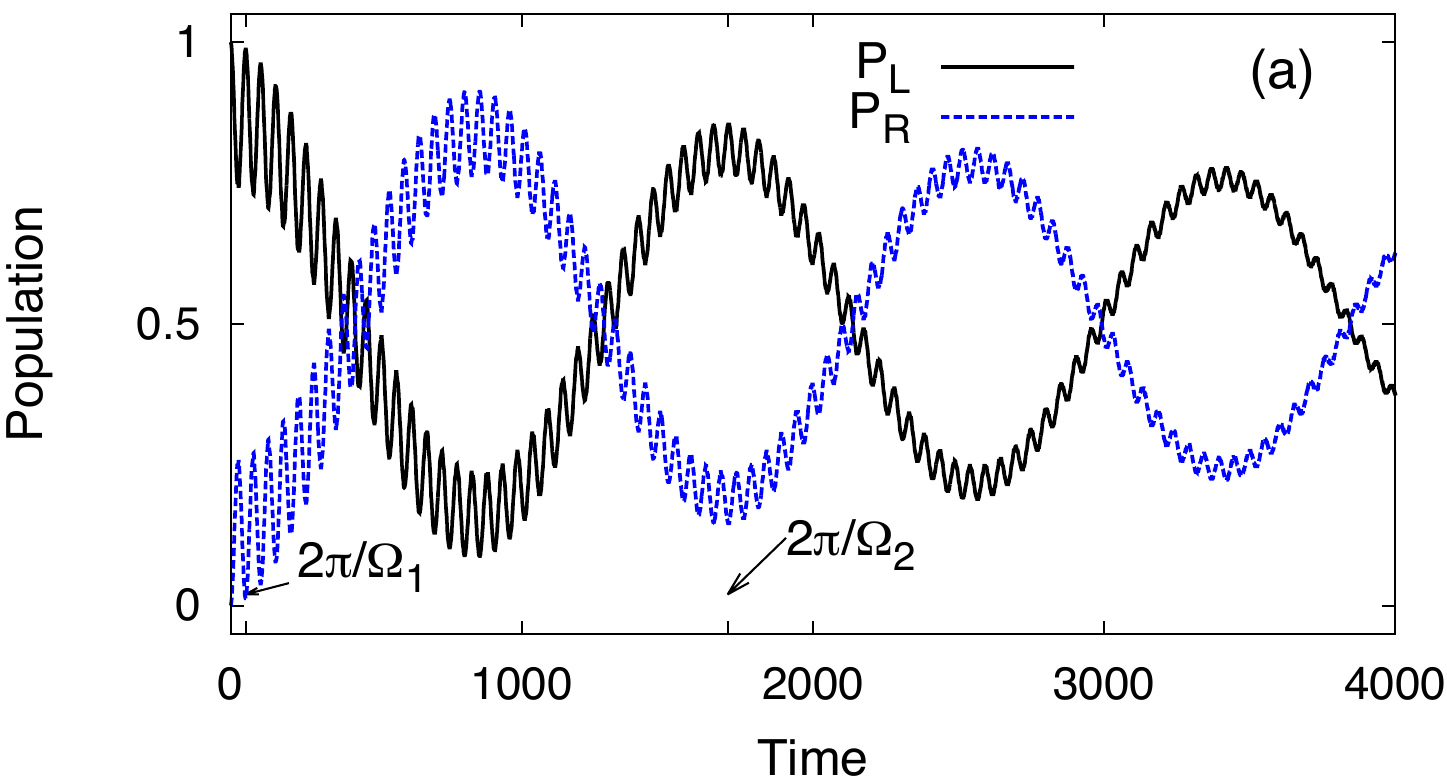}
\includegraphics[width=8.5cm ,angle=0]{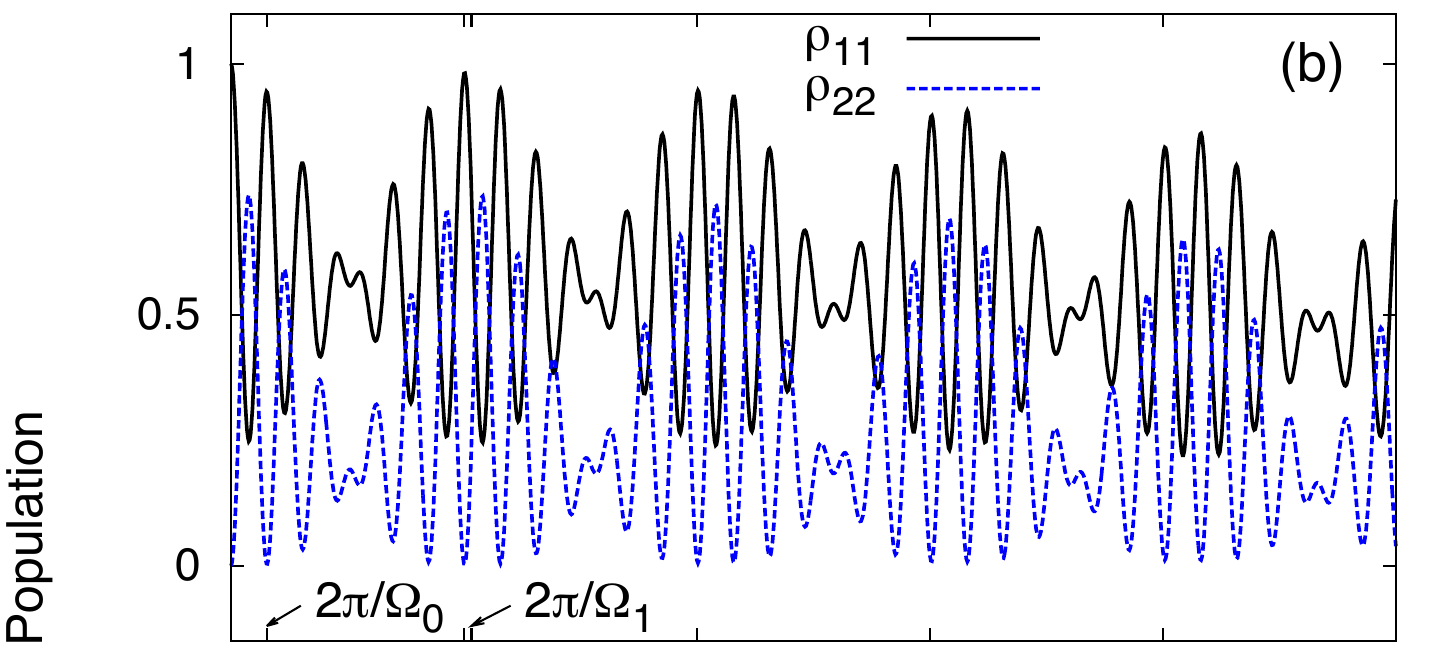}
\includegraphics[width=8.5cm ,angle=0]{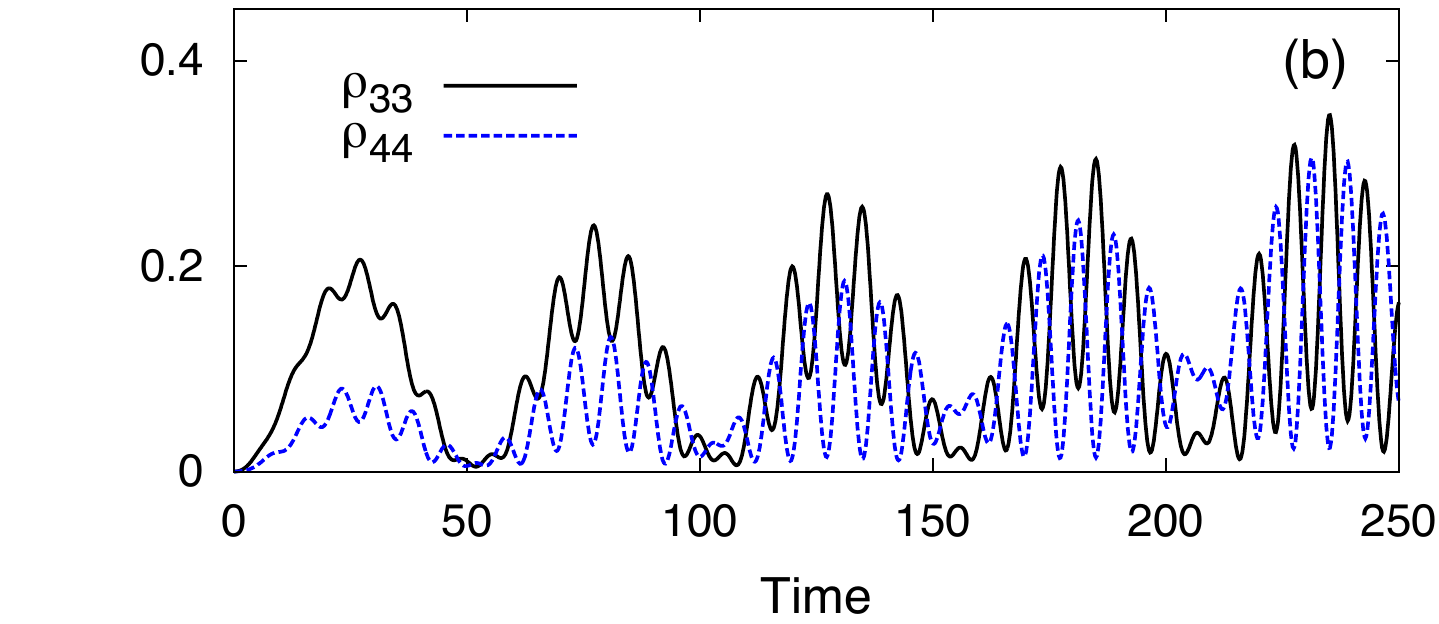}
\caption{\small{(Color online) Time evolution of the populations for $\gamma=0.0004~\omega_{0}$ and $T=0.001~\hbar\omega_{0}/k_{B}$, corresponding to phase point $1$ of the phase diagram in Fig.~\ref{fig9}. ($a$) -  Left-  and right-well populations $P_{L/R}=\rho_{11/33}+\rho_{22/44}$ vs time. ($b$) - Populations $\rho_{ii}$ of the states $|Q_{i}\rangle$ vs time. The results are obtained from the Bloch-Redfield master equation~(\ref{BR_ME-2}). The frequencies $\Omega_{i}$ are defined in Sec.~\ref{DDS}. Time is in units of $\omega_{0}^{-1}$.}}
\label{fig10}
\end{center}
\end{figure} 
The Ohmic bath spectral density function $J(\omega)=M\gamma\omega\exp(-\omega/\omega_{c})$, with cutoff frequency $\omega_{c}=50~\omega_{0}$, is assumed.
The potential is the same as in Fig.~\ref{fig2}, with $\epsilon=0$ and $\Delta U=1.4~\hbar\omega_{0}$. In what follows $t_{0}=0$ and the initial condition is  $\rho(0)=|Q_{1}\rangle\langle Q_{1}|$, i.e., $\rho_{kk}(0)=\delta_{k1}$.
\subsection{Dynamics of the double-doublet system}
\label{DDS-dynamics}
In this section we show the time evolution of the populations $\rho_{kk}$, the diagonal elements of the reduced density matrix in the localized basis $|Q_{1}\rangle,\dots,|Q_{4}\rangle$  (discrete variable representation), at the phase points denoted by diamonds in Fig. \ref{fig9}.\\
\begin{figure}[htbp]
\begin{center}
\includegraphics[width=8.5cm ,angle=0]{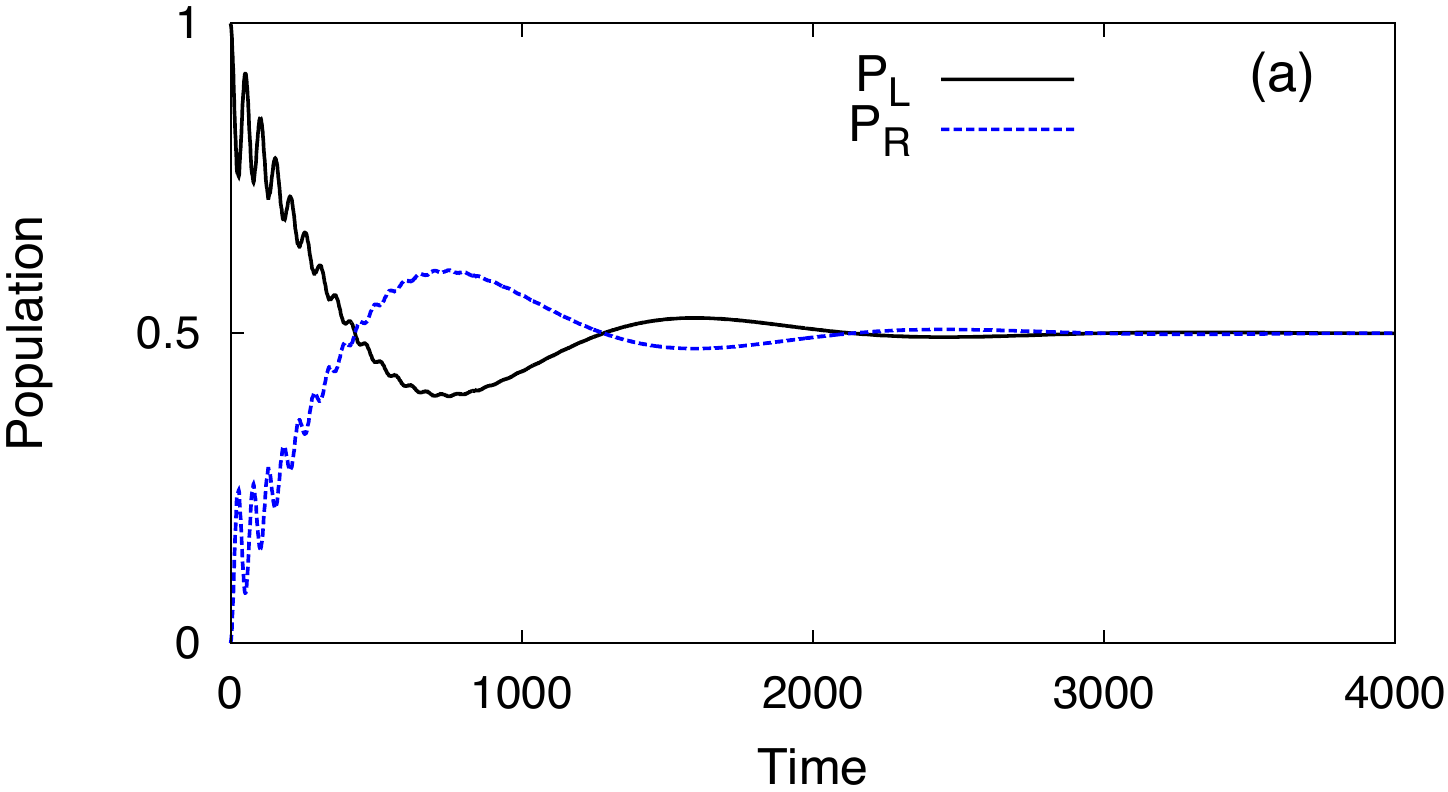}
\includegraphics[width=8.5cm ,angle=0]{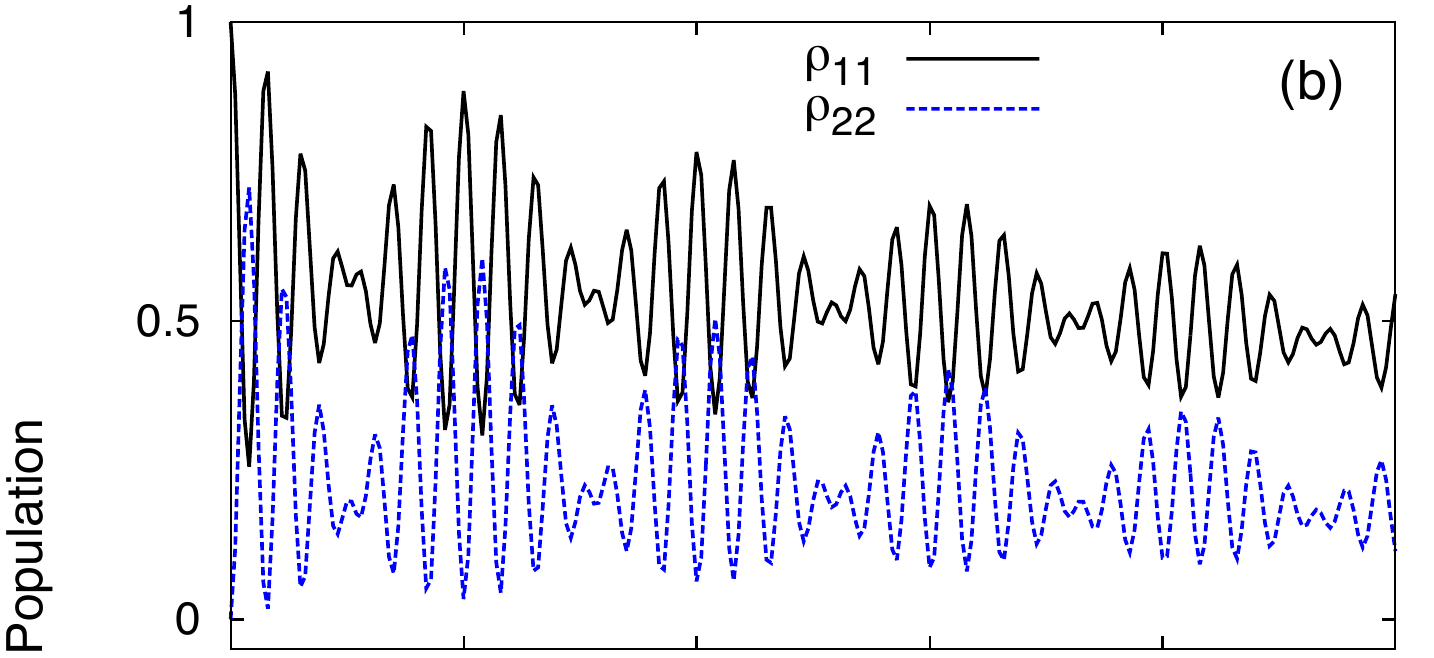}
\includegraphics[width=8.5cm ,angle=0]{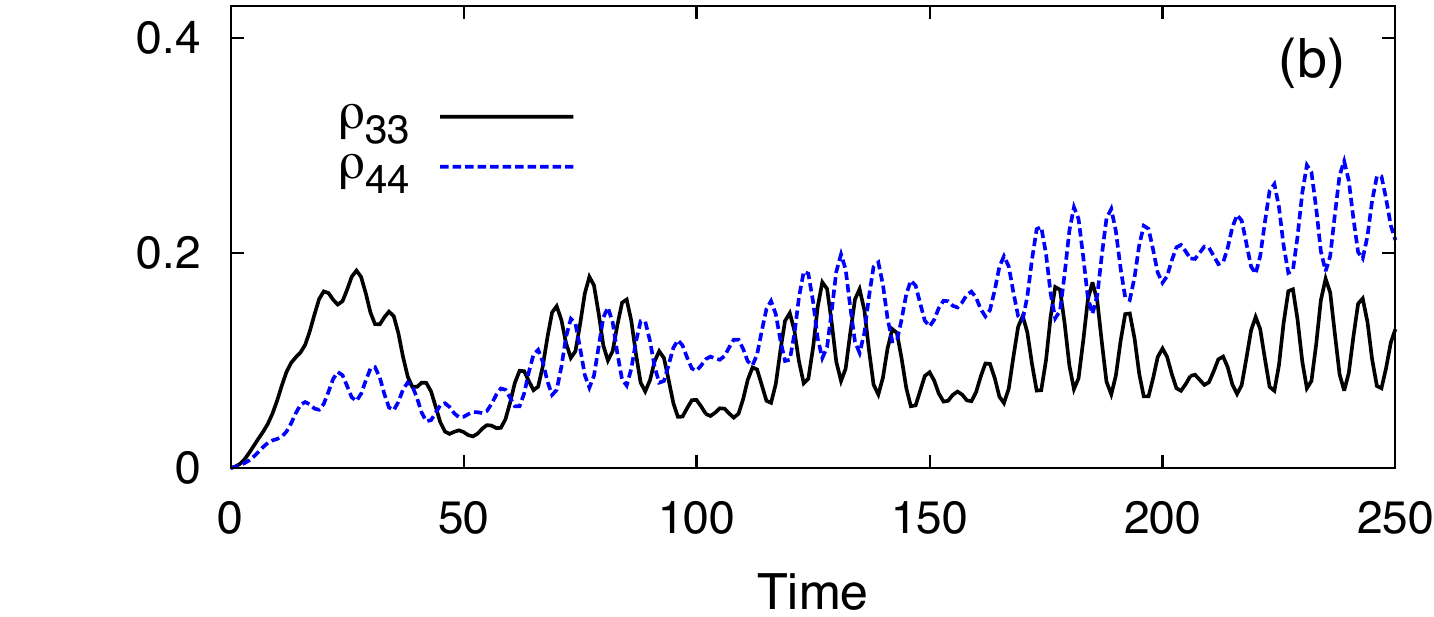}
\caption{\small{(Color online) Same as in Fig.~\ref{fig10},  but at stronger coupling and higher temperature: $\gamma=0.005~\omega_{0}$ and $T=0.005~\hbar\omega_{0}/k_{B}$ (phase point $2$ of the phase diagram in Fig.~\ref{fig9}). The relaxation is much faster than at phase point $1$. Time is in units of $\omega_{0}^{-1}$.}}
\label{fig11}
\end{center}
\end{figure}
\begin{figure}[htbp]
\begin{center}
\includegraphics[width=8.5cm ,angle=0]{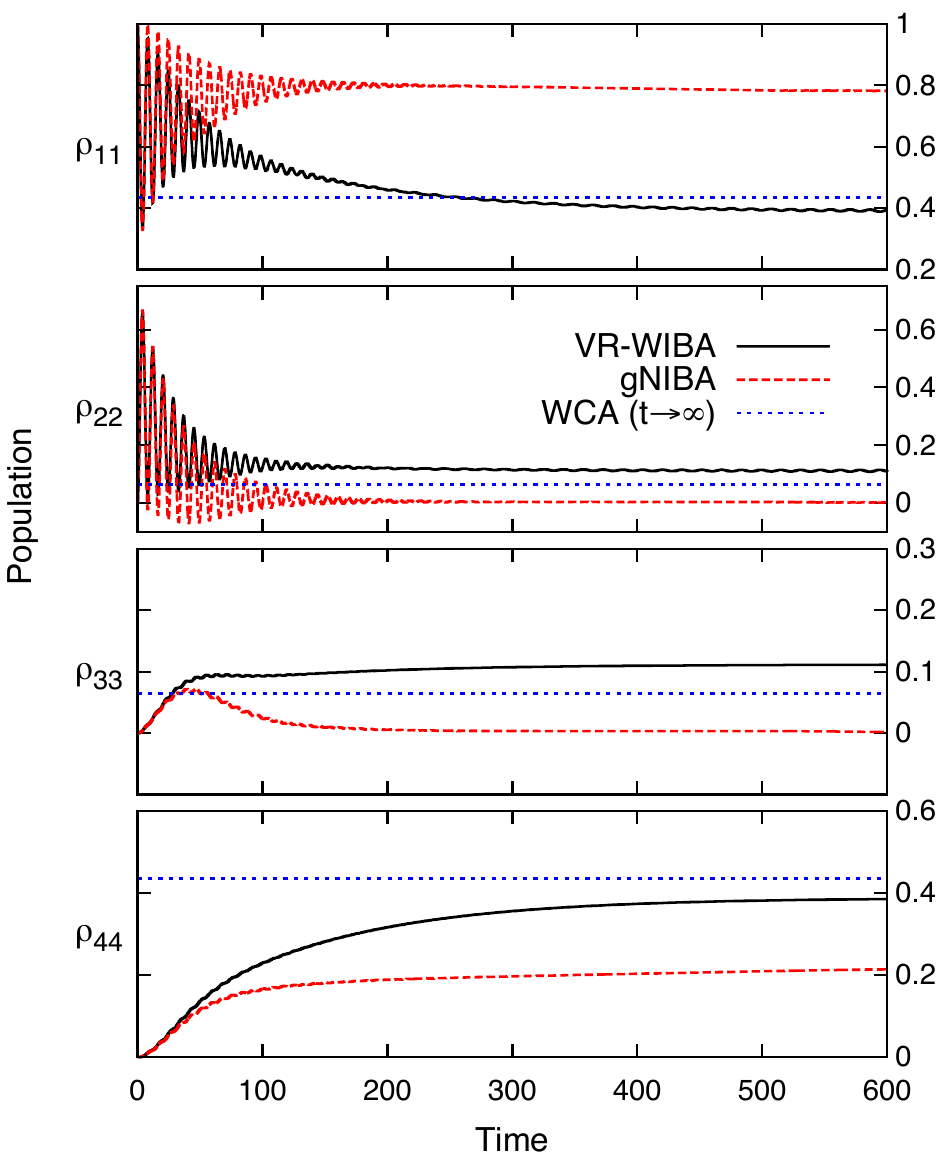}
\caption{\small{(Color online) Time evolution of the populations of the states $|Q_{i}\rangle$ in the crossover regime for $\gamma=0.05~\omega_{0}$ and $T=0.1~\hbar\omega_{0}/k_{B}$, corresponding to phase point $3$ of the phase diagram in Fig.~\ref{fig9}. Comparison between VR-WIBA and gNIBA results. Dotted blue lines are the equilibrium values given by Eq.~(\ref{rho_asympt}) in the weak coupling approximation (WCA). Time is in units of $\omega_{0}^{-1}$.}}
\label{fig12}
\end{center}
\end{figure}
\indent The results in the dynamical regime $A$ of complete coherence, at phase points $1$ and $2$, are obtained by solving the Bloch-Redfield master equation~(\ref{BR_ME-2}). Details are in Appendix~\ref{bloch-redfield}. Phase points $3$-$8$ are in the crossover $C$ and incoherent $D$ regimes, out of the reach of the perturbative Bolch-Redfield approach. The dynamics in these regimes is thus evaluated by numerical integration of the generalized master equation~(\ref{GME}) within our novel scheme, the VR-WIBA, and, for comparison, within the gNIBA.\\
\begin{figure}[htbp]
\begin{center}
\includegraphics[width=8.5cm ,angle=0]{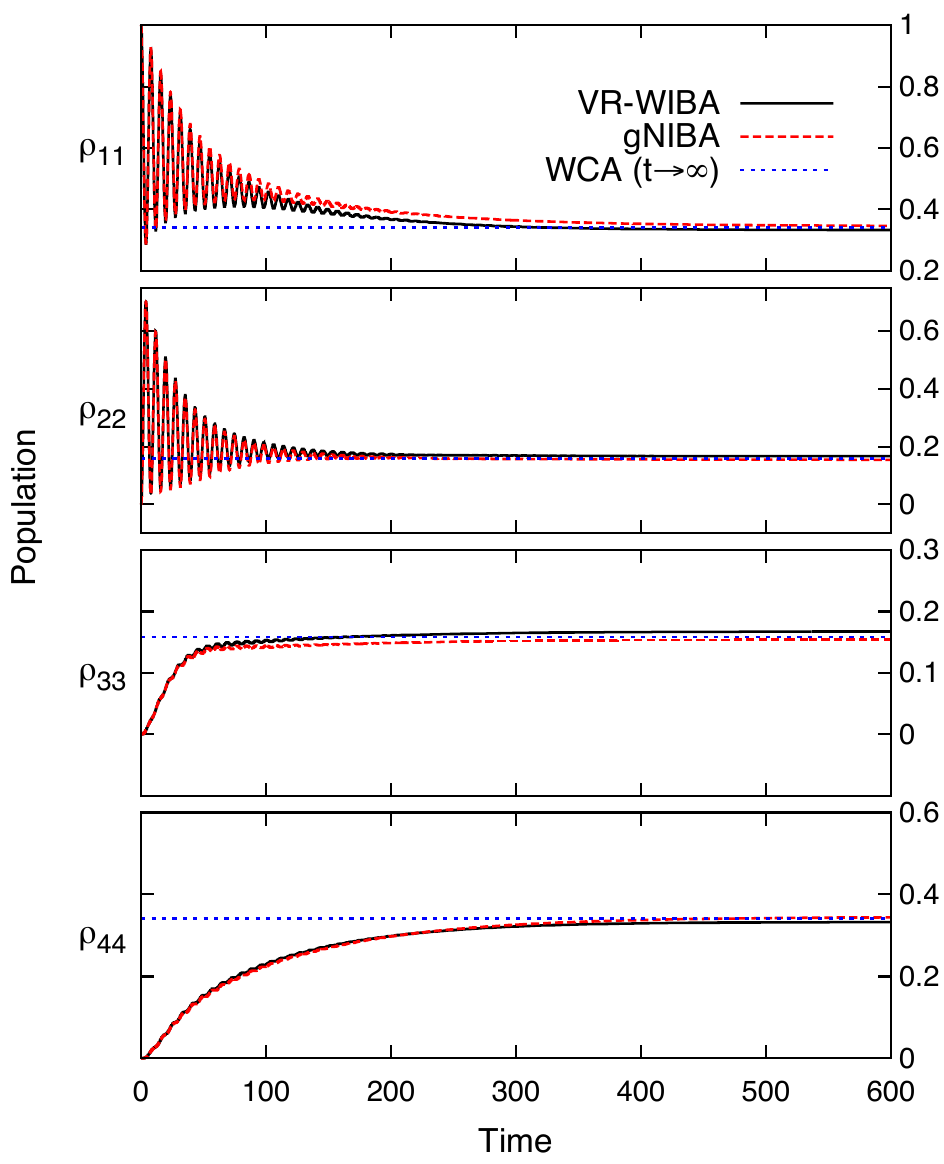}
\caption{\small{(Color online) Same as in Fig.~\ref{fig12} but at phase point $4$ of the phase diagram: $\gamma=0.02~\omega_{0}$ and $T=0.5~\hbar\omega_{0}/k_{B}$. In this regime a good agreement is found between the VR-WIBA and gNIBA results. Time is in units of $\omega_{0}^{-1}$.}}
\label{fig13}
\end{center}
\end{figure}
\indent In Fig.~\ref{fig10} the time evolution of  left- and right-well populations $P_{L/R}=\rho_{11/33}+\rho_{22/44}$ and of individual populations is shown at phase point $1$, at very low temperature and weak coupling, well within the validity of the Bloch-Redfield approach. The dynamics of  $P_{L/R}$  displays a weakly damped oscillatory behavior of frequency $\Omega_{2}$ with the small oscillations of frequency $\Omega_1$ featured also in the free dynamics (see Fig.~\ref{fig3}).
The short-time behavior of the individual populations is shown in Fig.~\ref{fig10}(b): As in the free case, fast oscillations of frequency $\Omega_0$ around an oscillatory envelope of frequency $\Omega_1$ are present. We remark that the fast intrawell oscillations found at this very low temperature are a result of the initial condition involving both the energy doublets (see Sec.~\ref{free DDS}).\\
\indent The second time evolution of the double-doublet system is shown in Fig.~\ref{fig11} and corresponds to phase point $2$ of the diagram in the same dynamical regime as that of Fig.~\ref{fig10}. Oscillations of left- and right-well populations are damped out after a few periods. Individual populations  show the same qualitative features as for phase point $1$. However, the damping of tunneling oscillations, which constitute the envelope of the fast intrawell motion, is now visible on the time scale set by $\Omega_{1}$.\\
\indent Phase point $3$ is in the crossover regime, at weak coupling and intermediate to low temperature, with respect to $\Omega_0$ (see Fig.~\ref{fig9}).  In this dissipation regime perturbative approaches, such as the Bloch-Redfield, fail. The dynamics obtained by using the path-integral approach shows coherence only at the level of intrawell motion while the tunneling is incoherent, as shown in Fig.~\ref{fig12}. We find that VR-WIBA and gNIBA give different predictions. On the basis of the discussion in Sec. \ref{phase diagram} it is  expected that the latter scheme fails, as phase point $3$ is outside its  validity domain. The reason is that the temperature is not sufficiently high to justify the NIBA for the biased intrawell effective two-level systems $\{|Q_{1}\rangle,|Q_{2}\rangle\}$ and $\{|Q_{3}\rangle,|Q_{4}\rangle\}$. The weak-coupling predictions of Eq.~(\ref{rho_asympt}) for the asymptotic values of the populations, are also shown in Fig.~\ref{fig12}, as well as in Figs.~\ref{fig13}, for comparison.\\
\begin{figure}[htbp]
\begin{center}
\includegraphics[width=8.5cm ,angle=0]{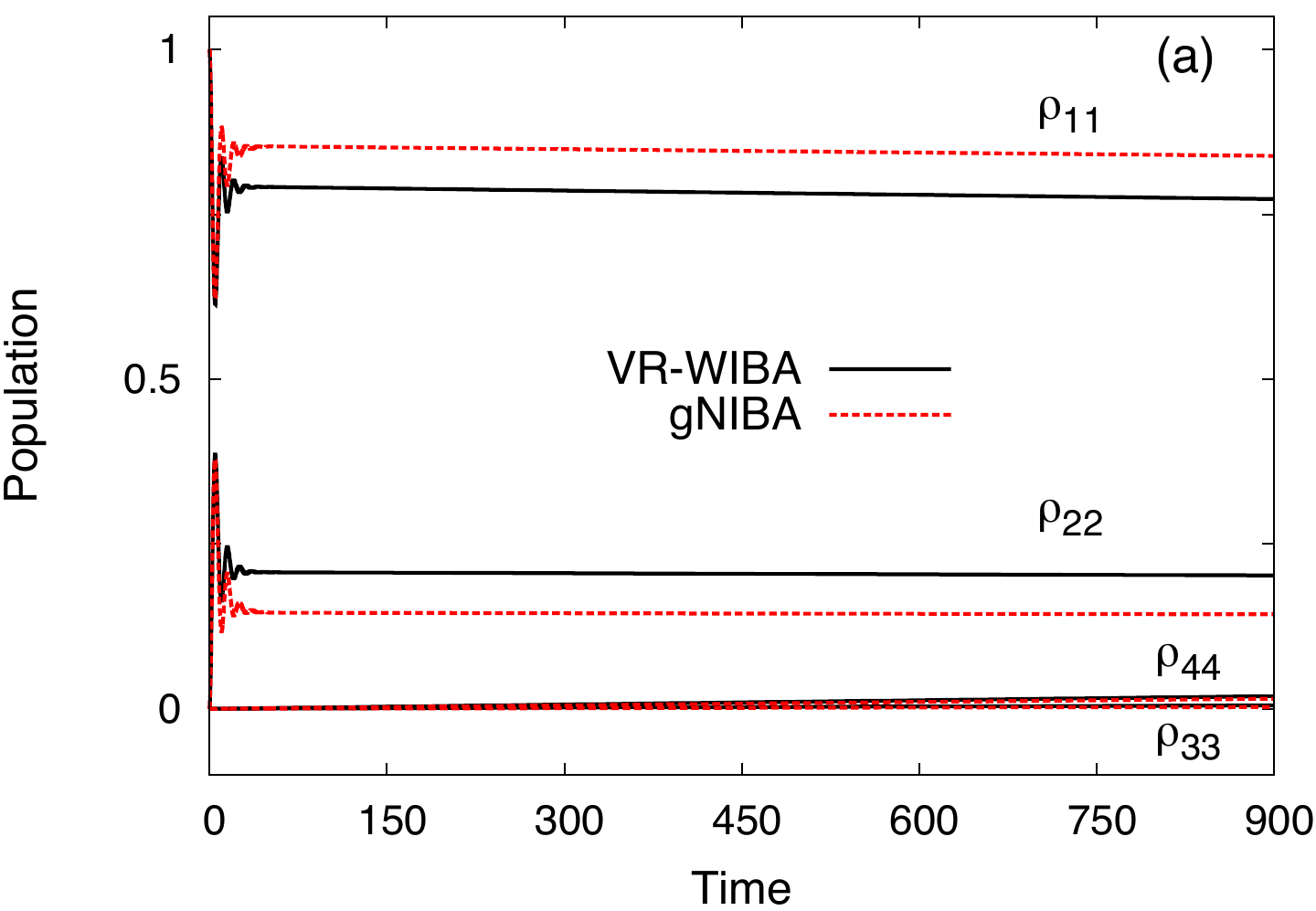}\\
\includegraphics[width=8.5cm ,angle=0]{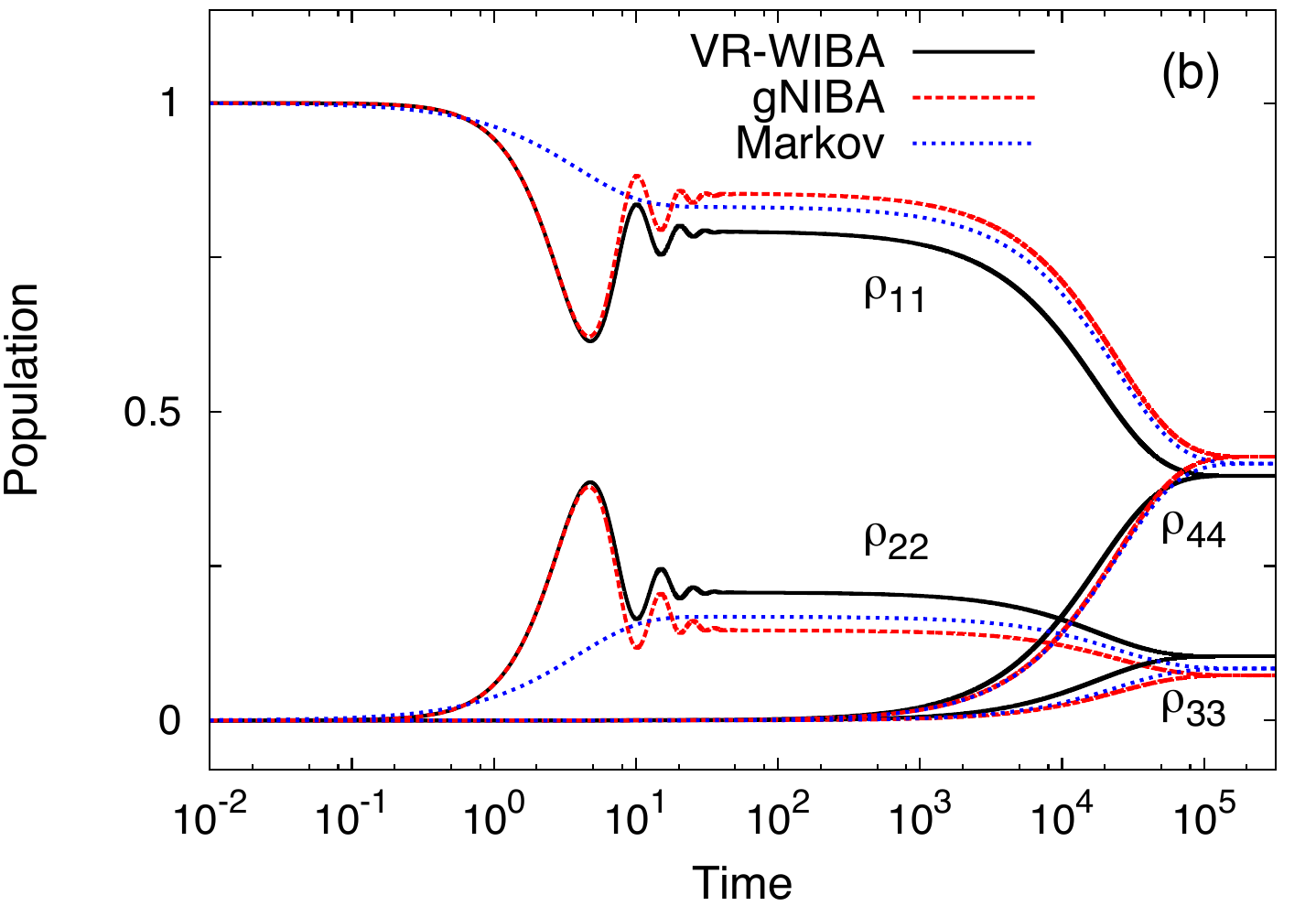}
\caption{\small{(Color online) Time evolution of the populations of the states $|Q_{i}\rangle$ in the crossover regime for $\gamma=0.25~\omega_{0}$ and $T=0.25~\hbar\omega_{0}/k_{B}$, corresponding to phase point $5$ of the phase diagram in Fig.~\ref{fig9}. Comparison between the results of VR-WIBA and gNIBA. 
($a$) - Short time dynamics.  ($b$) - Dynamics up to equilibrium with time in log scale. The predictions of the Markov approximated gNIBA master equation~(\ref{GME-markov}) are shown for comparison. Time is in units of $\omega_{0}^{-1}$.}}
\label{fig14}
\end{center}
\end{figure}
\indent The next time evolution is provided at phase point $4$ of the diagram of Fig.~\ref{fig9}, namely at weak coupling and high temperature with respect to the intrawell frequency $\Omega_{0}$. Contrary to the previous case, here both the VR-WIBA and the gNIBA are expected to give the correct prediction. Indeed,  as shown in Fig. \ref{fig13}, the results obtained within the two approximation schemes coincide and the asymptotic values of the populations reproduce the weak-coupling predictions.\\ 
\begin{figure}[htbp]
\begin{center}
\includegraphics[width=8.5cm ,angle=0]{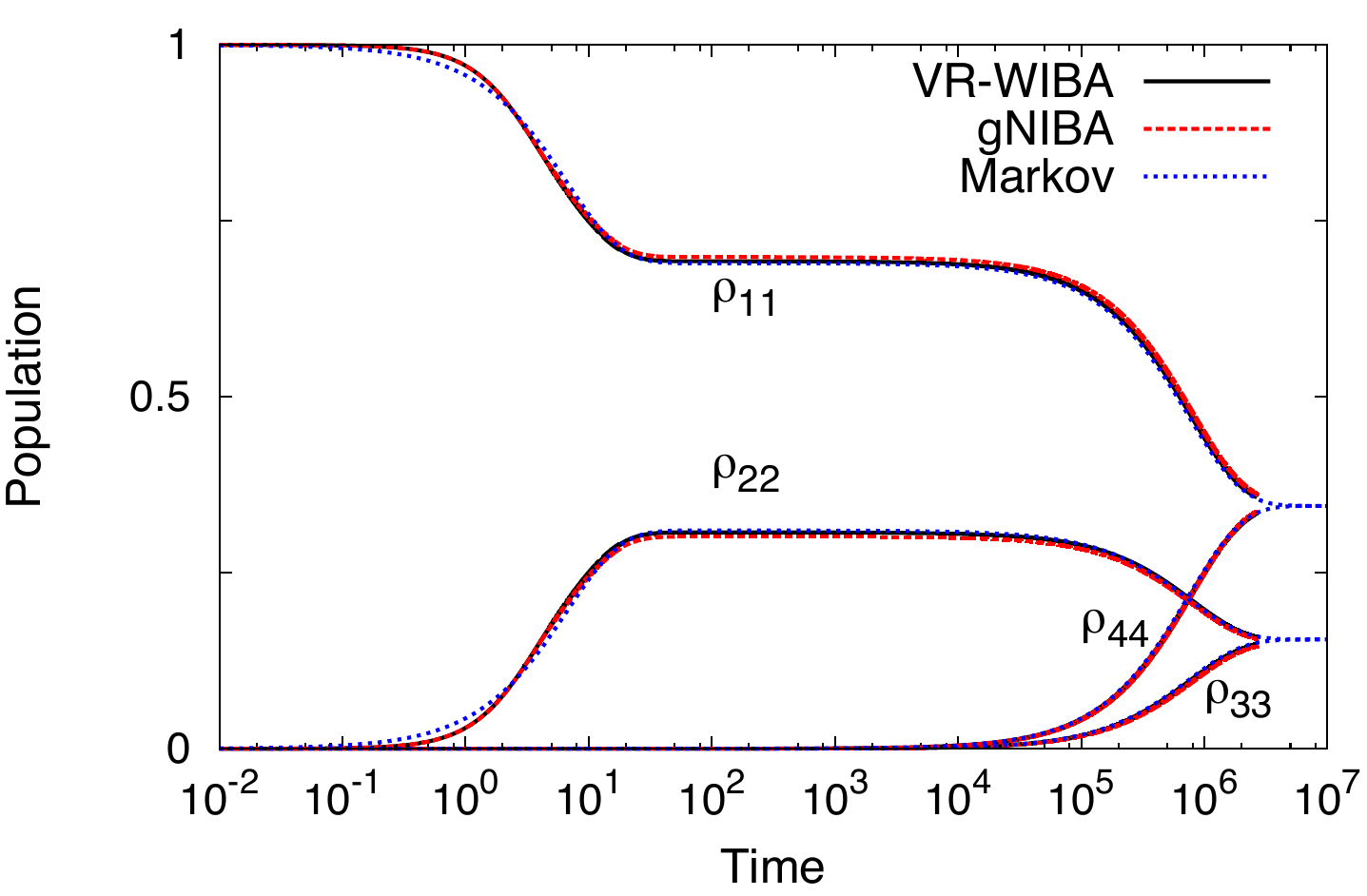}
\caption{\small{(Color online) Time evolution of the populations of the states $|Q_{i}\rangle$ in the incoherent regime for  $\gamma=0.5~\omega_{0}$ and $T=0.5~\hbar\omega_{0}/k_{B}$, corresponding to phase point $6$ of the phase diagram. Contrary to phase point $5$, no oscillations are present in the transient. Time is in units of $\omega_{0}^{-1}$.}}
\label{fig15}
\end{center}
\end{figure} 
\indent The fifth time evolution is at phase point $5$, in the crossover regime $C$, and displays  strongly damped intrawell oscillations and slow incoherent tunneling relaxation.  Again, the gNIBA predictions differ from those of the VR-WIBA, which confirms that, also in this coupling regime, the gNIBA is suited only for high temperatures.\\
\indent  As shown in Fig.~\ref{fig14}, at phase point $5$, which is characterized by a large value of the coupling strength, the equilibrium is reached on a very large time scale as compared to the results in the same crossover regime but at weaker coupling. Moreover, as shown in Fig.~\ref{fig14}(b), the dynamics features a transient metastable configuration which eventually decays to the equilibrium configuration. This feature is due to the nonequilibrium initial condition highlighting the different time scales involved. In particular, the first decay is towards a sort of \emph{intrawell} equilibrium which further decays, due to the tunneling, to the \emph{true} equilibrium. In  Fig.~\ref{fig14}(b) is also shown, for comparison, the solution of the Markov-approximated gNIBA master equation~(\ref{GME-markov}): It gives a good estimate for the relaxation time  at phase point $5$, even if it does not capture the oscillations in the transient dynamics.\\
\indent We note that, even if taking the inter-blip interactions to the first order in $\gamma$ is questionable at strong coupling, a comparison of the WIBA with numerically exact techniques (QUAPI \cite{Makarov1995}) for an asymmetric TLS at strong coupling, suggests that the WIBA attains a good performance~\cite{NesiPRB2007}. Since in our scheme, the VR-WIBA, the intrawell motion is  treated according to the WIBA, we expect to obtain reliable predictions  for the double-doublet system as well.\\
\indent The last phase point in the phase diagram of Fig.~\ref{fig9} is in the incoherent regime $D$. 
In Fig. \ref{fig15} we show the time evolution at phase point $6$, which is characterized by strong coupling and high temperature. The VR-WIBA and gNIBA predictions for the incoherent relaxation of the populations agree and coincide with  the solution of the Markov approximated gNIBA master equation~(\ref{GME-markov}), which is also shown for comparison~\cite{Thorwart2001}. As for phase point $5$, also in this case the transient dynamics features a metastable configuration due to the nonequilibrium initial condition. Moreover, due to the strong coupling, the relaxation to equilibrium is reached on a very large time scale.\\
\indent The picture that emerges from the various dynamical examples shown in this section is that at weak dissipation and low temperature the equilibrium is reached at long times, because the system is close to be isolated from the environment (see Figs.~\ref{fig10} and~\ref{fig11}). Large relaxation times are found also in the opposite regime of strong dissipation, where the \emph{viscosity} of the environment slows down the relaxation to equilibrium, as shown in Figs.~\ref{fig14} and~\ref{fig15}. In the intermediate situation (Figs.~\ref{fig12} and~\ref{fig13}) the equilibrium is reached on the shortest time scale.\\
\section{Conclusions}
\label{conclusions}
In this work we give a comprehensive account of the dissipative dynamics of the double-doublet system in Ohmic environment. This is done by the combined use of Born-Markov master equation and path-integral approaches, the latter within the proposed novel approximation scheme called  \emph{weakly interacting VR-blip approximation} (VR-WIBA).\\ 
\indent  This scheme takes into account, at the level of the intrawell dynamics, the time nonlocal correlations featured in the Feynman-Vernon influence functional. As a result, the VR-WIBA contains and extends the domain of validity of the preexisting generalized non-interacting blip approximation (gNIBA), and succeeds in describing the crossover dynamical regime occurring at intermediate temperatures in a broad range of coupling strengths. The crossover regime, which is to a large extent inaccessible to previous approximation schemes, is characterized by coherence in the intrawell motion and incoherent tunneling dynamics.\\ 
\indent  At weak coupling and low temperature we use the Bloch-Redfield master equation to account for the coherent oscillatory behavior of the intrawell and  tunneling dynamics. A weak coupling approach is also used to check the VR-WIBA predictions for the stationary configuration. This is done for values of the  coupling which are small with respect to the intrawell characteristic frequency, even if  the tunneling dynamics is strongly damped.\\ 
\indent The combined use of master equation and path-integral techniques  accounts for the dissipative dynamics of the double-doublet system in a large region of the parameter space where a four-state truncation of the Hilbert space is justified. To show this, we  establish a phase diagram which describes the dynamics corresponding to the various dissipation regimes and the domains of validity of the techniques used in this work.\\
\indent The dissipative dynamics of  the double-doublet system is obtained in each of the accessible dissipation regimes, ranging from very weak coupling/low temperature to strong coupling/high temperature. This is done by using a perturbative Bolch-Redfield master equation technique and the nonperturbative path-integral approach within our novel approximation scheme (VR-WIBA). A comparison with the gNIBA results is made.\\ 
\indent  Some final remarks are in order. Even if the calculations presented in this work are performed for an unbiased double-well potential, the applicability of the VR-WIBA is not limited to the symmetric case but comprises also the asymmetrical one. Indeed a static bias can be taken into account provided that  intra- and interwell dynamics occur on very different time scales, i.e., for inter-doublet energy separation much larger than the intra-doublet one~\cite{Magazzu2015}. This condition is not very restrictive as it is fulfilled for any double-well potential with two energy doublets below the top of the barrier.\\ 
\indent Second, the generalization of the VR-WIBA to broadband sub-Ohmic or super-Ohmic environments is possible, although care must be taken in establishing, from time to time, the validity of the approximations discussed throughout this work.\\
\indent Finally, the approximation of a Hilbert space truncated to the first few energy states is at the basis of the description in terms of localized states given by the discrete variable representation (DVR). The approximation is valid as long as the temperature is not high enough to involve higher lying energy levels. Within this restriction, the discrete variable and energy representations are equivalent, since they are related by a unitary transformation.\\
\indent Path-integral developments based on the Feynman-Vernon influence functional are carried out in the position representation and allow for  the exact elimination of the heat bath in the Caldeira-Leggett model. This is due to the fact that, since the interaction is mediated by the position operator, the action is split into a bare system and a system-bath term, as shown in Sec.~\ref{PI approach}. In the approximate treatment with a restricted Hilbert space, the DVR is the natural choice, as in this representation the position operator is diagonal. Moreover, the resulting picture of interacting charges allows for approximations in the opposite  limits of strong and weak coupling~\cite{Weiss2012}, and even for dealing with intermediate situations, as done in the present work. The DVR basis, which was first introduced by A. J. Leggett \emph{et al.}~\cite{Leggett1987} to deal with the spin-boson model in the strong dissipation regime, is also the more convenient in numerically exact \emph{ab initio} calculations~\cite{Makri1995}. This is because it provides an unequally spaced spatial grid with a minimum number of grid points, chosen in a physically sensible fashion. Another typical path-integral approach is that of  coherent-state path-integral~\cite{Klauder1985,Zhang1990,Inomata1992,Burghardt1998,Burant2002,Novikov2004,Kleinert2009,Wilson2011}. 
For example, path-integral evaluations of the propagator using coherent states  have been done by using quantum Monte Carlo~\cite{Zhang2003} and stochastic~\cite{Koch2008} techniques. Another numerical approach for quantum dynamics, based on a discretized coherent state representation, has been introduced in Ref.~\cite{Andersson2001}.
Nevertheless, to the best of our knowledge, no analytical real-time coherent-state path-integral technique for double well potentials has been developed. Moreover, the formulation of the coherent-state path-integral, widely and routinely used in many areas of physics, despite its success, gives rise to unsolved mathematical problems~\cite{Wilson2011}.
\section*{Acknowledgments}
We acknowledge financial support from the Collaborative Research Project SFB 631. This work was partially supported by MIUR through Grant. No. PON$02\_00355\_3391233$, 
Tecnologie per l'ENERGia e l'Efficienza energETICa - ENERGETIC.
\appendix
\section{Bloch-Redfield master equation}
\label{bloch-redfield} 
The energy representation of the double-doublet system is given by the four energy eigenstates $|E_{n}\rangle$ satisfying
\begin{equation}
\hat{H}_S|E_{n}\rangle=\hbar\omega_{n}|E_{n}\rangle \qquad\qquad (n=1,\dots,4).
\end{equation}
We define 
\begin{equation}
\omega_{nm}=\omega_{n}-\omega_{m}\qquad\text{and}\qquad q_{nm}=\langle E_{n}|\hat{q}|E_{m}\rangle. 
\end{equation}
In the energy representation, to first order in the coupling and under the assumption that the memory time of the bath is short compared to the characteristic times in the evolution of the density matrix (Markov approximation), the following \emph{Bloch-Redfield master equation} can be derived~\cite{Blum2012} from the microscopical model given in Sec.~\ref{model}:
\begin{equation}\label{BR_ME}
\dot{\rho}_{nm}^{E}(t)=-i\omega_{nm}\rho_{nm}^{E}(t)+\sum_{k,l}\mathcal{L}_{nm,kl}\rho_{kl}^{E}(t).
\end{equation}
The Bloch-Redfield tensor is 
\begin{equation}\label{BlochRedfieldTensor}
\begin{aligned}
\mathcal{L}_{nm,kl}&=q_{nk}\left(O_{lm}+P_{lm}\right)+q_{lm}\left(O_{nk}-P_{nk}\right)\\
&-\sum_{j}\left[\delta_{kn}q_{jm}\left(O_{lj}+P_{lj}\right)+\delta_{lm}q_{nj}\left(O_{jk}-P_{jk}\right)\right],
\end{aligned}
\end{equation}
where  
\begin{equation}
\begin{aligned}
O_{nm}=q_{nm}\int_{0}^{\infty}d\tau&\int_{0}^{\infty}d\omega\frac{J(\omega)}{\pi\hbar}\\
&\times\coth\left(\frac{\beta\hbar\omega}{2}\right)\cos(\omega\tau) e^{-i \omega_{nm}\tau}
\end{aligned}
\end{equation}
and
\begin{equation}
P_{nm}=q_{nm}\omega_{nm}\int_{0}^{\infty}d\tau\int_{0}^{\infty}d\omega\frac{J(\omega)}{\omega\pi\hbar}\cos(\omega\tau) e^{-i \omega_{nm}\tau}.
\end{equation}
To perform the integral over $\tau$ we use
\begin{equation}\label{principal-value}
\int_{0}^{\infty}d\tau e^{i\tilde{\omega}\tau}=\pi\delta(\tilde{\omega})+i\mathcal{P}\frac{1}{\tilde{\omega}}.
\end{equation}
Neglecting the principal value, which gives a frequency shift, $O_{nm}$ and $P_{nm}$ read
\begin{equation}\label{Qnm}
O_{nm}=q_{nm}\frac{J(|\omega_{nm}|)}{2\hbar}\coth\left(\frac{\beta\hbar|\omega_{nm}|}{2}\right)
\end{equation}
and
\begin{equation}\label{Pnm}
P_{nm}=\frac{q_{nm}\omega_{nm}}{2\hbar}\frac{J(|\omega_{nm}|)}{|\omega_{nm}|}.
\end{equation}
For $\omega_{nm}>0$
\begin{equation}
\begin{aligned}
O_{nm}-P_{nm}=&~O_{mn}+P_{mn}\\
=&~q_{nm}\frac{J(|\omega_{nm}|)}{\hbar}n_{\beta}(\omega_{nm}),
\end{aligned}
\end{equation}
while for $\omega_{nm}<0$
\begin{equation}
\begin{aligned}
O_{nm}-P_{nm}=&~O_{mn}+P_{mn}\\
=&~q_{nm}\frac{J(|\omega_{nm}|)}{\hbar}\left( n_{\beta}(|\omega_{nm}|)+1\right).
\end{aligned}
\end{equation}
Here $n_{\beta}(\omega_{nm})$ is the expectation value of the number of  bath excitations of energy $\hbar\omega_{nm}$ at temperature $T=(k_{B}\beta)^{-1}$.
\subsection{Analytic solution in the full secular approximation}
Setting $\rho_{nm}^{E}(t)=e^{-i\omega_{nm}(t-t_{0})}\sigma_{nm}(t)$, Eq.~(\ref{BR_ME}) becomes 
\begin{equation}\label{ME_sigma}
\dot{\sigma}_{nm}(t)=\sum_{kl}\mathcal{L}_{nm,kl}\Omega_{nm,kl}(t)\sigma_{kl}(t),
\end{equation}
where  $\Omega_{nm,kl}(t)=\exp\left[i(\omega_{nm}-\omega_{kl})(t-t_{0})\right]$. We have $\sigma(t_{0})=\rho(t_{0})$.
\\ 
\indent  The \emph{full secular approximation} (FSA) consists in neglecting the terms in the master equation for which $\omega_{\mu\nu}-\omega_{\kappa\lambda}\neq 0$. Mathematically this condition reads 
\begin{equation}
\Omega_{\mu\nu\kappa\lambda}(t)\rightarrow 
\left(\delta_{\kappa\mu}\delta_{\lambda\nu}+\delta_{\kappa\lambda}\delta_{\mu\nu} \right)\Omega_{\mu\nu\kappa\lambda}(t).
\end{equation}
In the FSA the equations for diagonal elements decouple from those for non-diagonal elements of $\sigma$.\\ 
\indent  Specifically, the dynamics of $\sigma(t)$ is given by a master equation for the diagonal elements and a set of independent equations for the 
non-diagonal elements.
The master equation for the diagonal elements reads
\begin{equation}\label{ME_diag_sigma}
\dot{\sigma}_{nn}(t)=\sum_{k}\mathcal{L}_{nn,kk}\sigma_{kk}(t),
\end{equation}
where, for $n\neq k$,  
\begin{equation}
\mathcal{L}_{nn,kk}=q_{nk}\left(O_{kn}+P_{km}\right)+q_{kn}\left(O_{nk}-P_{nk}\right)
\end{equation}
and $\mathcal{L}_{n,n}=-\sum_{k}\mathcal{L}_{k,n}$.\\
\indent The solution of Eq.~(\ref{ME_diag_sigma}) is
\begin{equation}
\sigma_{nn}(t)=\sum_{ij}S_{ni}e^{\lambda_{i}(t-t_{0})}(S^{-1})_{ij}\sigma_{jj}(t_{0}),
\end{equation}
where $S$ is the transformation that diagonalizes the matrix  $L_{nk}=\mathcal{L}_{nn,kk}$ with eigenvalues $\lambda_{i}$.
 From the definition of $\sigma(t)$ we have $\rho_{nn}^{E}(t)=\sigma_{nn}(t)$.\\ 
\indent  The uncoupled equations for the non-diagonal elements of $\sigma(t)$ are
\begin{equation}\label{nondiag_sigma}
\dot{\sigma}_{nm}(t)=-\mathcal{L}_{nm,nm}\sigma_{nm}(t),
\end{equation}
with
\begin{equation}
\begin{aligned}
\mathcal{L}_{nm,nm}=&(q_{nn}-q_{mm})\left[O_{nn}-P_{nn}-(O_{mm}+P_{mm})\right]\\
&+\sum_{j\neq m}q_{jm}(O_{mj}+P_{mj})+\sum_{j\neq n}q_{nj}(O_{jn}-P_{jn}).
\end{aligned}
\end{equation}
In our specific problem, due to the symmetry of the potential, the diagonal matrix elements $q_{ii}$ of  the position operator in the energy representation vanish.\\ 
\indent  The solutions of Eq.~(\ref{nondiag_sigma}) are
\begin{equation}
 \sigma_{nm}(t)=e^{-\mathcal{L}_{nm,nm}(t-t_{0})}\sigma_{nm}(t_{0}),
\end{equation}
so that the non-diagonal elements of the density matrix in the energy representation are
\begin{equation}
 \rho_{nm}^{E}(t)=e^{-i\omega_{nm}(t-t_{0})}e^{-\mathcal{L}_{nm,nm}(t-t_{0})}\rho_{nm}^{E}(t_{0}).
\end{equation}
\indent Once the solution for $\rho$ in the energy basis is known, to pass to the localized basis $\{|Q_{j}\rangle\}$ we  perform the transformation
\begin{equation}
 \rho^{DVR}_{nm}(t)=\sum_{ij}T_{ni}\rho_{ij}^{E}(t)T_{jm}^{\dag},
\end{equation}
where $T_{ij}=\langle E_{i} | q_{j}\rangle$.
\section{Parameters}
\label{parameters}
\indent Here we give the list of transition amplitudes per unit time and bias factors for the symmetric double-doublet system considered in this work. The two indexes in $\Delta_{ij}$ and $\epsilon_{ij}$ specify the states $q$ and $q'$. In terms of the characteristic frequencies $\Omega_{0}$, defined in Eq.~(\ref{Omega0}), $\Omega_{1}=(E_{4}-E_{3})/\hbar$, and  $\Omega_{2}=(E_{2}-E_{1})/\hbar$, the coefficients $\Delta_{ij}=\langle Q_{i}|\hat{H}_{S}|Q_{j}\rangle$, introduced in Eq.~(\ref{delta-j}), are  
\begin{equation}\label{Delta-DDS}
\Delta_{12}=\Delta_{21}=\Delta_{43}=\Delta_{34}=v^2u\Omega_0,\\
\end{equation}
and
\begin{equation}\label{Delta-DDS2}
\left\{ 
  \begin{array}{l l}
&\Delta_{13}=\Delta_{31}=\Delta_{24}=\Delta_{42}=v^2u(\Omega_{1}-\Omega_{2})/2,\\
&\Delta_{23}=\Delta_{32}=v^2(\Omega_{1}+u^2\Omega_{2})/2,\\
&\Delta_{14}=\Delta_{41}=v^2(u^2\Omega_{1}+\omega_{2})/2.
  \end{array} \right.
\end{equation}
The constant $u$ depends on the parameters of the potential ($u\simeq 0.585$ in our problem) and $v=(1+u^{2})^{-1/2}$.\\
\indent The biases $\epsilon_{ij}=(\langle Q_{i}|\hat{H}_{S}| Q_{i}\rangle-\langle Q_{j}|\hat{H}_{S}| Q_{j}\rangle)/\hbar$  (see Eq.~(\ref{epsilon-j})) are
\begin{equation}\label{epsilon-DDS}
\begin{aligned}
 \epsilon_{12}&=\epsilon_{13}=\epsilon_{43}=\epsilon_{42}=-\epsilon_{21}=-\epsilon_{41}=-\epsilon_{34}=-\epsilon_{24}\\
 &=v^2(u^2-1)\Omega_0,\\
 \epsilon_{14}&=\epsilon_{41}=\epsilon_{23}=\epsilon_{32}=0.
 \end{aligned}
\end{equation}
Because of inequality~(\ref{Inequality}), the amplitudes per unit time of the vibrational relaxation transitions (Eq.~(\ref{Delta-DDS})) are greater than those of tunneling transitions   (Eq.~(\ref{Delta-DDS2})).\\
\indent Note that the relations among the parameters in Eqs.~(\ref{Delta-DDS})-(\ref{epsilon-DDS})  reflect the symmetry of the potential considered in this work and do not hold for a biased bistable potential.\\ 
\section{Propagator in Laplace space}
\label{propagator-laplace}
First we give the expressions for the blip times $\tau$ and the sojourn times $\sigma$ 
\begin{equation}
\begin{aligned}
&\tau_j=t_{2j}-t_{2j-1}\\
&\sigma_j=t_{2j+1}-t_{2j}.\\
\end{aligned}
\end{equation}
The approximation on the paths made in Sec.~\ref{path-selection} implies that, if $\rho(t_{0})=|Q_{j}\rangle\langle Q_{j}|$, then each path contributing to the population $\rho_{kk}(t)$ has an even number $2n$ of transitions.  Consider the series of integrals 
\begin{equation}\label{integrals}
\int_{t_{0}}^t dt_{2n}\int_{t_{0}}^{t_{2n}}dt_{2n-1}\dots\int_{t_{0}}^{t_3}  dt_2\int_{t_{0}}^{t_2}  dt_1.
\end{equation}
By using repeatedly the rule 
\begin{equation}\label{exchange_rule}
\int_{t_{0}}^{t_{j+1}}dt_{j}\int_{t_{0}}^{t_{j}}dt_{j-1}=\int_{t_{0}}^{t_{j+1}}dt_{j-1}\int_{t_{j-1}}^{t_{j+1}}dt_{j},
\end{equation}
Eq.~(\ref{integrals}) can be put into the form
\begin{equation}\label{integrals2}
\begin{aligned}
\int_{t_{0}}^t dt_1 &\int_{t_1}^t dt_2\int_{t_2}^t dt_3\dots\int_{t_{2n-2}}^t dt_{2n-1}\int_{t_{2n-1}}^t dt_{2n}\\
=&\int_0^{\bar{t}} d\sigma_0\int_0^{\bar{t}-\sigma_0} d\tau_1\int_0^{\bar{t}-\tau_1-\sigma_0} d\sigma_1\dots\\
&\times\int_0^{\bar{t}-\dots-\tau_{n-1}} d\sigma_{n-1}\int_0^{\bar{t}-\dots-\sigma_{n-1}} d\tau_n .\\
\end{aligned}
\end{equation}
Notice that there is no integration over the last sojourn time, since it is fixed by the length of the interval $\bar{t}=t-t_{0}$.\\ 
\indent  By using repeatedly the rule $\int_{0}^{\infty}dt\int_{0}^{t}dt'=\int_{0}^{\infty}dt'\int_{t'}^{\infty}dt$ and the relation $\bar{t}=\sigma_{n}+\tau_{n}+\dots+\tau_{1}+\sigma_{0}$, the Laplace transform of Eq.~(\ref{integrals2}) reads 
\begin{equation}\label{Laplace-integrals}
\begin{aligned}
&\int_0^{\infty} d\bar{t} e^{-\lambda \bar{t}}\int_0^{\bar{t}} d\sigma_0\dots\int_0^{\bar{t}-\dots-\sigma_{n-1}} d\tau_n\\
&=\int_0^{\infty} d\sigma_n e^{-\lambda \sigma_n}\int_0^{\infty}d\tau_n e^{-\lambda \tau_n}\dots \int_0^{\infty} d\sigma_{0} e^{-\lambda \sigma_{0}}.
\end{aligned} 
\end{equation}
\indent Consider the time integrals in Eq.~(\ref{Laplace-integrals}) of an  amplitude $A$ corresponding to a path with $2n$ transitions distributed into  $N$ subpaths. Since the amplitude $A_{j}$ of the $j$-th subpath doesn't depend on the initial sojourn time $\sigma_{0}^{j}$, Eq.~(\ref{Laplace-integrals}) can be cast  in the form
\begin{equation}\label{Laplace-integrals2}
\begin{aligned}
\frac{1}{\lambda}\prod_{j=1}^{N}
\int_0^{\infty}d\tau_{k_{j}}^{j} e^{-\lambda \tau_{k_{j}}^{j}}\dots\int_0^{\infty}d\tau_1^{j} e^{-\lambda \tau_1^{j}}\int_0^{\infty}d\sigma_0^{j} e^{-\lambda \sigma_0^{j}},\\
\end{aligned} 
\end{equation}
where the factor $1/\lambda$ derives from integration over $\sigma_{n}$ in Eq.~(\ref{Laplace-integrals}). Equation~(\ref{rho_lambda_DDS}) follows from equality of Eqs.~(\ref{Laplace-integrals})  and~(\ref{Laplace-integrals2}). The function $\hat{g}_{q_{j},q_{j+1}}$ reads
\begin{equation}
\begin{aligned}
\hat{g}_{q_{j},q_{j+1}}&(\lambda)=\sum_{k_{j}=1}^{\infty}\sum_{\text{paths}_{2k_{j}}}\int_0^{\infty}d\tau_{k_{j}}^{j} e^{-\lambda \tau_{k_{j}}^{j}}\\
&\dots\int_0^{\infty}d\tau_1^{j} e^{-\lambda \tau_1^{j}}\int_0^{\infty}d\sigma_0^{j} e^{-\lambda \sigma_0^{j}}A_{j}(\tau_{1}^{j},\sigma_{1}^{j},\dots,\tau_{k_{j}}^{j}).
\end{aligned} 
\end{equation}
The Laplace transform of the propagator  in Eq.~(\ref{prop-TLS}) of the two-level system corresponding to the $j$-th subpath is 
\begin{equation}
\begin{aligned}
\hat{G}_{q_{j}q_{j+1}}(\lambda)=
\sum_{k_{j}=1}^{\infty}\int_{0}^{\infty}\mathcal{D}_{k_{j},\lambda}A_{j}(\tau_{1}^{j},\sigma_{1}^{j},\dots,\tau_{k_{j}}^{j}),
\end{aligned} 
\end{equation}
where 
\begin{equation}
\begin{aligned}
\int_{0}^{\infty}\mathcal{D}_{k,\lambda}=&\sum_{\text{paths}_{2k}}\int_0^{\infty} d\sigma_{k} e^{-\lambda \sigma_{k}}\int_0^{\infty}d\tau_{k} e^{-\lambda \tau_{k}}\\
&\times\dots \int_0^{\infty} d\sigma_{0} e^{-\lambda \sigma_{0}}.
\end{aligned} 
\end{equation}
\indent Since integration over the last sojourn time $\sigma_{k}$ yields a $1/\lambda$ factor, we have
\begin{equation}
\lambda\hat{G}_{q_{j}q_{j+1}}(\lambda)=\hat{g}_{q_{j}q_{j+1}}(\lambda).
\end{equation}
\section{Propagator in terms of irreducible kernels}
\label{TLSpropagator}
 Consider $\vec{\rho}(t)$, the two-dimensional population vector of a two-level system and assume that it satisfies the following generalized master equation
\begin{equation}\label{TLS-GME}
\dot{\vec{\rho}}(t)=\int_{t_{0}}^{t}dt'\mathcal{K}(t-t')\vec{\rho}(t'),
\end{equation}
where $\mathcal{K}$ is the matrix of the so-called  irreducible kernels.
\indent In Laplace space Eq.~(\ref{TLS-GME}) reads 
\begin{equation}\label{TLS-GME-l}
\vec{\rho}(\lambda)=\frac{1}{\lambda}\left[\hat{\mathcal{K}}(\lambda)\vec{\rho}(\lambda)+\vec{\rho}(t_{0})\right].
\end{equation}
Iterating Eq.~(\ref{TLS-GME-l}) we get 
\begin{equation}
\vec{\rho}(\lambda)=\frac{1}{\lambda}\sum_{n=0}^{\infty}\left[\frac{\hat{\mathcal{K}}(\lambda)}{\lambda}\right]^{n}\vec{\rho}(t_{0}).
\end{equation}
Since $\vec{\rho}(t)=\mathcal{G}(t,t_{0})\vec{\rho}(t_{0})$, where $\mathcal{G}$ is the matrix whose elements are the propagators $\mathcal{G}_{fi}=G(q_{f},q_{f},t;q_{i},q_{i},t_{0})$, we have
\begin{equation}
\lambda\hat{\mathcal{G}}(\lambda)=\sum_{n=0}^{\infty}\left[\frac{\hat{\mathcal{K}}(\lambda)}{\lambda}\right]^{n},
\end{equation}
which is Eq.~(\ref{propagatorTLS-laplace}) in vector notation.
\section{VR-WIBA kernels}
\label{vrwiba-kernels} 
If $Q_{k}$ and $Q_{j}$ belong to different wells, then the populations of the states $|Q_{k}\rangle$ and $|Q_{j}\rangle$ in the VR-WIBA generalized master equation~(\ref{GME}) are connected by the NIBA kernels
\begin{equation}\label{SDD_NIBA_kern}
\begin{aligned}
K_{kj}^{N}(t)=2\Delta_{kj}^2e^{-q_{kj}^{2}S(t)}\cos{\left(\epsilon_{kj}t+q_{kj}^{2}R(t)\right)},
\end{aligned}
\end{equation}
where
 \begin{equation}
 \begin{aligned}
 &\Delta_{kj}=\frac{1}{\hbar}\langle Q_{k}|\hat{H}_{S}|Q_{j}\rangle\\ 
 &\epsilon_{kj}=\frac{1}{\hbar}\left(\langle Q_{k}| \hat{H}_{S}|Q_{k}\rangle-\langle Q_{j}| \hat{H}_{S}|Q_{j}\rangle\right)\\
 & q_{kj}^{2}=(Q_{k}-Q_{j})^{2}.
 \end{aligned}
 \end{equation}
\indent  If $Q_{k}$ and $Q_{j}$ belong to the same well, then the kernel is  the WIBA kernel: $K_{kj}^{W}(t)=K_{kj}^N(t)+K_{kj}^{BN}(t)$. The beyond-NIBA correction is~\cite{NesiPRB2007}
\begin{equation}
\begin{aligned}
K_{kj}^{BN}(t)=&8\Delta_{kj}^4\int_0^{t}d\tau\int_{0}^{t-\tau}d \tau' e^{- q_{kj}^{2}S(\tau)- q_{kj}^{2}S(\tau')}\\
\times&\sin(\epsilon_{kj} \tau')\cos(q_{kj}^{2}R(\tau'))p_{kj}(t-\tau-\tau')\\
\times& [q_{kj}^{2}X(t,\tau')\cos(\epsilon_{kj}\tau+q_{kj}^{2}R(\tau))\\
-& q_{kj}^{2}\Lambda(t,\tau',\tau)\sin(\epsilon_{kj}\tau+ q_{kj}^{2}R(\tau)) ],
\end{aligned}
\end{equation}
where 
\begin{equation}
\Lambda(t,\tau',\tau)=S(t)+S(t-\tau'-\tau)-S(t-\tau)-S(t-\tau')
\end{equation}
and
\begin{equation}
 X(t,\tau')= R(t)-R(t-\tau').
\end{equation}
In the calculations $S$ and $R$ are taken in the scaling limit form given in Eq.~(\ref{Q_t}).\\ 
\indent  The functions $p_{kj}$ obey the equations 
 \begin{equation}
\dot{p}_{kj}(t)=\int_0^t dt' K_{kj}^{N,(+)}(t-t')p_{kj}(t')
\end{equation}
with initial condition $p_{kj}(0)=1$ and kernel
\begin{equation}
K_{kj}^{N,(+)}(t)=-4\Delta_{kj}^2 e^{- q_{kj}^{2}S(t)}\cos(\epsilon_{kj}\tau)\cos( q_{kj}^{2}R(t)).
\end{equation}
Note that, by the symmetry of the problem, the four functions $p_{12},p_{21},p_{34}$, and $p_{43}$ are the same for the symmetric double-doublet system.
\bibliographystyle{apsrev4-1}
%\bibliography{bibliography}
%

\end{document}